\documentclass[runningheads,a4paper]{llncs}

\usepackage[american]{babel}

\usepackage[%
rm={oldstyle=false,proportional=true},%
sf={oldstyle=false,proportional=true},%
tt={oldstyle=false,proportional=true,variable=true},%
qt=false%
]{cfr-lm}
\usepackage{graphicx}
\usepackage{bbm}
\usepackage{amsmath}
\usepackage{amssymb}
\usepackage{latexsym}

\usepackage{subfigure}
\usepackage{bm}

\usepackage{paralist}

\usepackage{cite}

\usepackage[T1]{fontenc}

\usepackage[math]{blindtext}

\usepackage{csquotes}

\usepackage{microtype}

\usepackage{url}
\makeatletter
\def\Url@twoslashes{\mathchar`\/\@ifnextchar/{\kern-.2em}{}}
\g@addto@macro\UrlSpecials{\do\/{\Url@twoslashes}}
\makeatother
\makeatletter
\g@addto@macro{\UrlBreaks}{\UrlOrds}
\makeatother

\usepackage{xcolor}

\usepackage[
bookmarks=false,
breaklinks=true,
colorlinks=true,
linkcolor=black,
citecolor=black,
urlcolor=black,
pdfpagelayout=SinglePage,
pdfstartview=Fit
]{hyperref}
\usepackage[all]{hypcap}

\usepackage{pdfcomment}

\usepackage[capitalise,nameinlink]{cleveref}
\crefname{section}{Sect.}{Sect.}
\Crefname{section}{Section}{Sections}

\usepackage{xspace}

\DeclareFontFamily{U}{MnSymbolC}{}
\DeclareSymbolFont{MnSyC}{U}{MnSymbolC}{m}{n}
\DeclareFontShape{U}{MnSymbolC}{m}{n}{
    <-6>  MnSymbolC5
   <6-7>  MnSymbolC6
   <7-8>  MnSymbolC7
   <8-9>  MnSymbolC8
   <9-10> MnSymbolC9
  <10-12> MnSymbolC10
  <12->   MnSymbolC12%
}{}
\DeclareMathSymbol{\powerset}{\mathord}{MnSyC}{180}

\begin{document}

\input glyphtounicode.tex
\pdfgentounicode=1

\title{Multi-scale Community Detection in Temporal Networks Using Spectral Graph Wavelets}
\titlerunning{Multi-scale Community Detection in Temporal Networks}

\author{Zhana Kuncheva\inst{1,2} \and Giovanni Montana\inst{2,3} }

\institute{
Clinical Development and Mathematics, C$4$X Discovery Ltd, M1 $3$LD, UK.\and
Department of Mathematics, Imperial College London, SW$7$ $2$AZ, UK.\\
\email{z.kuncheva12@imperial.ac.uk}\and
Department of Biomedical Engineering, King's College London, SE$1$ $7$EH, UK.\\
\email{giovanni.montana@kcl.ac.uk}
}
			
\maketitle

\begin{abstract}

Spectral graph wavelets introduce a notion of scale in networks, and are thus used to obtain a local view of the network from each node. By carefully constructing a wavelet filter function for these wavelets, a multi-scale community detection method for monoplex networks has already been developed. This construction takes advantage of the partitioning properties of the network Laplacian. 
In this paper we elaborate on a novel method which uses spectral graph wavelets to detect multi-scale communities in temporal networks. To do this we extend the definition of spectral graph wavelets to temporal networks by
adopting a multilayer framework. We use arguments from Perturbation Theory to investigate the
spectral properties of the supra-Laplacian matrix for clustering purposes in temporal networks. Using 
 these properties, we construct a new wavelet filter function, which attenuates the influence of uninformative eigenvalues and centres the filter around
eigenvalues which contain information on the coarsest description
of prevalent community structures over time. We use the spectral graph wavelets as feature vectors in a connectivity-constrained clustering
procedure to detect multi-scale communities at different scales, and refer to this method as Temporal Multi-Scale Community Detection (TMSCD). We validate the performance of TMSCD and a competing methodology on various benchmarks. The advantage of TMSCD is the automated selection of relevant scales at which communities should be sought. \\
\end{abstract}


\section{Introduction}

Networks are used to model complex relationships in a wide range of real-life applications throughout the social, biological, physical, information technology
and engineering sciences. Due to limitations in data collection and storage, mainly static (monoplex) networks have been studied. However, many real-world systems have relationships between entities that evolve over time~\cite{Holme2012}. Technological advances have increased the amount of recorded temporal information. As a result, 
the sequence of networks describing changes occuring over time have been formalized as \textit{temporal networks}
(also known as time-varying or time-stamped)~\cite{Holme2012}. Examples of temporal networks include the functional brain networks ~\cite{BETZEL2014,Bassett2011}, social media interactions~\cite{Zhao2012}, financial markets~\cite{Aynaud2010} or politics~\cite{Macon2012} . 

One aspect of temporal network analysis is the discovery of community structures, which are groups of nodes that are more densely connected to each other than they are to the rest of the network~\cite{Newman2006}. Changes in the configuration of communities over time signals important turns in the evolution of the system. Real data networks are often observed to have communities with a hierarchical structure referred to as multi-scale communities~\cite{Sales-Pardo2007}. Changes in the community structure over time might take place either at one scale or across all scales of the community structure. For this reason, there is interest in methods that are able to investigate communities at different
\textquotedblleft scales\textquotedblright over time~\cite{Newman2006,Tremblay2014,Mucha}.

Some recent approaches to community detection in temporal networks rely on a simple network aggregation procedure whereby
all time networks are first collapsed into a single network. Afterward traditional algorithms for community detection can be used
~\cite{Traud2011}. These methods, however, ignore valuable information 
about the evolution of the community structures over time. Other methods 
investigate each time network individually~\cite{Aynaud2010,Fenn2009,Macon2012,JasonD.Lee}, thus ignoring the dependence of community structures between neighbouring time points.

There exist methods that extend algorithms from one to multiple networks by using the multilayer formulation of a temporal network~\cite{Kivel}. This special data structure allows inter-layer couplings between neighbouring time networks~\cite{Kivel,Mucha}. One method, which is extended in this way, is the \textit{modularity maximization}~\cite{Newman2006}. The modularity of a network is defined as the number of connections within a community compared to the expected number of such connections in an equivalent random network. The generalisation of the \textit{modularity maximization} introduced in~\cite{Mucha} overcomes the obstacles mentioned earlier by using the multilayer formulation. In this way it introduces a dependence between communities identified in one layer and connectivity patterns in other layers.

\textit{Modularity maximization} is controlled by a resolution parameter
$\gamma$, determining the size of the detected communities and supporting the detection of multi-scale communities. However, the range of parameter values must be manually selected. The importance of relevant scales is assessed using stability procedures, which compare the detected communities to those obtained from random networks~\cite{Schaub2012,Lambiotte2010}.
When dealing with real life data, communities at one or more scales can go undetected if appropriate parameter values are not
selected. The modularity maximization has also been used to investigate the effect of
constant inter-layer weights between consecutive time layers \cite{Bazzi2014} on the
behavior of community detection. 

Other approaches to multi-scale community mining in monoplex networks have been proposed to address some of the issues experienced by the modularity maximization. The method in~\cite{Tremblay2014} relies on spectral graph wavelets~\cite{Hammond2009} and introduces a notion of scale in the network. These wavelets are thus used to obtain a local view of the network from each node. The clustering properties of the spectrum of the Laplacian in clustering 
\cite{Fortunato2010,Bertrand2013,Luxburg} are used to construct a wavelet filter function which enables the spectral graph wavelets~\cite{Hammond2009} to be sensitive to multi-scale communities. 
Contrary to the modularity maximization, this method is able to automatically select the range of scales to be investigated for existing communities. For a better understanding of the current paper we suggest the reader gets acquainted with articles~\cite{Hammond2009,Tremblay2014}. 

In this paper we extend spectral graph wavelets to temporal networks. For this extension we use the supra-Laplacian of the temporal network, which is defined as the Laplacian of the matrix representation of its multilayer formulation. A challenge we face here is the need to 
take into account the fundamental difference between within-layer
and inter-layer edges when studying
the spectral properties of the supra-Laplacian~\cite{Kivel,Taylor2015,DeDomenico2013a}. Although some
studies explain the effect of different inter-layer weights over the eigenvalues of
the supra-Laplacian~\cite{Moreno2013,Sol2013}, there is no work related to the interpretation of the eigenvectors
of the supra-Laplacian for clustering purposes. 

Using Perturbation theory~\cite{Stewart1990,Bhatia1997}, we argue that the eigenvectors
corresponding to the smallest eigenvalues of the supra-Laplacian are a linear combination of the eigenvectors -- corresponding to all zero
eigenvalues -- of the Laplacian matrices of the separate time
layers. From spectral graph theory~\cite{Chung1996}, it is known that an eigenvector corresponding to the zero eigenvalue of the Laplacian matrix is not informative of the community structure. For this reason, the eigenvectors of the supra-Laplacian matrix, which can be obtained as approximations to these eigenvectors, cannot be used to identify communities within the time layers, and larger eigenvalues should be sought. Using the above arguments as a stepping stone, we propose a novel Temporal Multi-Scale Community Detection (TMSCD) method, which extends the notion of spectral graph wavelets to temporal networks and automatically selects relevant scales at which multi-scale community partitions are obtained. The method uses the relevance of a newly identified eigenvalue of the supra-Laplacian, which captures the coarse description of communities prevalent over time. 

In what follows, we first define the notation used throughout this paper in Section~\ref{sec:notation}. Section~\ref{sec:TMSCD} describes the method for multi-scale community detection in temporal networks which uses the spectral properties of the supra-Laplacian to identify relevant scale. In Section~\ref{sec:experiemnts} we compare the performance of TMSCD to the modularity maximization \cite{Mucha}. Section \ref{sec:conclusion} concludes this paper.

\section{Notation\label{sec:notation}}

Let $G=(V,A)$ be an $N$-node network where $V$ is the set of nodes and $A\in
\mathbb{R}
^{N\times N}$ is the adjacency matrix with edge weights between pairs of nodes $\left\{  A_{ij}|i,j\in\left\{
1,2,...,N\right\}  \right\}$. We only consider
undirected, adjacency matrices ($A_{ij}=A_{ji}$ for all
$i$ and $j$). The degree of a node $i$ is $d_{i}={\sum_{j=1}^{N}}A_{ij}$, and the degree matrix $D$ has $d_{1},d_{2},...,d_{N}$ on its main diagonal. Network $G$ is associated with the normalized Laplacian matrix $L=D^{-\frac{1}{2}}\left(D-A\right)D^{-\frac{1}{2}}$.

The networks representing different time points in the \textit{temporal network} are known as layers. We use the notation $G^{t}=\left(  V,A^{t}\right)  $ for layer $t$ in the \textit{temporal network} $\mathcal{T}=\left\{  G^{1},G^{2},...,G^{T}\right\}$, which is the ordered sequence of
networks for $t\in\left\{  1,2,...,T \right\}$ time points, and we denote node
$i$ in layer $t$ by $i_{t}$.  We work with \textit{temporal network} in which each node is present in
all layers. The multilayer framework of a \textit{temporal
network} considers a diagonal ordinal coupling of layers \cite{Kivel,Bassett2011,Mucha}. In essence, inter-layer weights exist only between corresponding nodes in neighboring time layers.   
We denote the inter-layer edge weight for node $i$ between consecutive layers $t$ and $t+1$ by $\omega_{i}^{t,t+1}\in\mathbb{R}$. Else
$\omega_{i}^{t,p}=0$ for $p\neq t-1,t+1$.

This \textit{temporal network} $\mathcal{T}$ has an associated adjacency matrix $\mathcal{A}$ of size $NT\times NT$ -- the supra-Adjacency matrix. The time-dependent diagonal blocks of $\mathcal{A}$, $\mathcal{A}_{t,t}$, are
the adjacency matrices $A^{t}$, and the off-diagonal blocks, $\mathcal{A}_{t,t+1}$, are the
inter-layer weight matrices $W^{t,t+1}=diag(\omega_{1}^{t,t+1},\omega_{2}%
^{t,t+1},...,\omega_{N}^{t,t+1})$. Else $\mathcal{A}_{t,p}$ is a
$N\times N$ zero matrix for $p\neq t-1,t+1$. 

The within-layer degree of node $i$ in layer $G^{t}$ is $d_{i}^{t}:={\sum_{j=1}^{N}}A_{ij}^{t}$
while the multilayer node degree of node $i$ in layer $G^{t}$ is $\mathfrak{d}_{i}^{t}:=d_{i}^{t}+\omega_{i}^{t,t-1}+\omega_{i}^{t,t+1}$. 
These define the degree matrix $\mathcal{D}$ with
diagonal entries $\mathcal{D}:=diag\left(  \mathfrak{d}_{1}^{1},\mathfrak{d}_{2}^{1}%
,...,\mathfrak{d}_{N}^{1},\mathfrak{d}_{1}^{2},...,\mathfrak{d}_{N}%
^{2},...,\mathfrak{d}_{N}^{T}\right)$. The \textit{normalized supra-Laplacian} $\mathcal{L}$ is computed as 
$\mathcal{L}\mathfrak{=}\mathcal{D}^{-\frac{1}{2}}\left(\mathcal{D-A}\right)\mathcal{D}^{-\frac{1}{2}}$.

\section{Temporal Multi-Scale Community Detection (TMSCD)\label{sec:TMSCD}}
The proposed TMSCD method is a multilayer
extension of the multi-scale community detection procedure via spectral graph wavelets developed in \cite{Tremblay2014}. The advantage of this method is the automated selection of relevant scales at which community partitions are obtained. In Section~\ref{sec:weight} we define new inter-layer weights at each node adapted for
community detection in temporal networks. In Section~\ref{sec:supWave} we extend the definition of a wavelet at a node (in a monoplex
network) to that for a wavelet at a node at a particular time layer. By
construction, a wavelet associated to a node at a time layer is local in the
whole \textit{temporal network}. The wavelet is centred around this node and spreads
on its neighbourhood, which consists of its neighbours in the current layer and
the corresponding nodes in the neighbouring time layers. The larger the
scale is, the larger the spanned neighbourhood is -- more nodes in current layer
and more nodes in neighbouring layers.

The most central part of our method is the construction of the wavelet filter function $g$. In Section~\ref{sec:specL} we use arguments from Perturbation theory to investigate the spectral properties of the supra-Laplacian matrix for community detection purposes, and propose a procedure for the selection of appropriate eigenvalues around which to center the wavelet filter function. In Section~\ref{sec:gBsplines}, we introduce the wavelet filter function based on a
$B$-spline, we choose the parameters of this function, and define relevant scales for community investigation.

Finally, in Section~\ref{sec:stability} for any given scale, we use the wavelet
of a node at a given time layer to cluster together nodes whose associated wavelets are correlated using an agglomerative connectivity-constrained
clustering procedure which respects the time sequence of the temporal
network. 

\subsection{Inter-Layer Couplings $\omega_{i}^{t,t+1}$
for Community Detection \label{sec:weight}}

The choice of inter-layer weights $\omega_{i}^{t,t+1}$ is
important - they control the ordinal coupling between time layers $t$ and
$t+1$ via the node $i$.  We believe that inter-layer couplings should be strong enough to indicate
similarity of a node's neighbourhood in two consecutive networks and indicate shared community structures over time. The main principle is, inter-layer weights should be strong enough to reflect on local topological similarity of nodes across layers, but they should not interfere with the within-layer structure.

Let the set of neighbours of node $i$ in layer $G^t$ be denoted by $\mathcal{N}_{i}^{t}:=\{j_{t}:A_{ij}^{t}=1\}.$

We introduce the inter-layer weight $\omega_{i\ }^{t,t+1}$ as
follows:%
\begin{equation}
\omega_{i}^{t,t+1}:=\frac{\left\vert \mathcal{N}_{i}^{t}\cap\mathcal{N}%
_{i}^{t+1}\right\vert }{2} . \label{eq:omegaz}%
\end{equation}

We refer to these inter-layer weights as LART-type since they were the basic ingredients of the LART algorithm  \cite{Kuncheva2015}. The LART algorithm is a method for the detection of communities that are shared by either some or all the layers in a multilayer network. The algorithm is based on a random walk and the transition probabilities defining the random walk are allowed to depend on the local topological similarity between layers at any given node.

It can be derived that $\omega_{i}^{t,t+1}\leq\frac{\min\left(  d_{i}^{t},d_{i}^{t+1}\right)}{2}$. From this follows that the multilayer node degree of $i_{t}$ is $\mathfrak{d}_{i}^{t}\leq2d_{i}^{t}$. 
Thus at least half of the influence, which node $i_{t}$ has over
the properties of $\mathcal{A}$ and therefore $\mathcal{L},$ comes
from the connections of node $i$ within layer $t,$ rather than from the
inter-layer weights $\omega_{i}^{t,t-1}$ and $\omega_{i}^{t,t+1}$.

\subsection{Construction of Spectral Graph Wavelets for Temporal
Networks\label{sec:supWave}}

Upon obtaining matrices $\mathcal{A}$ and $\mathcal{L}$, we construct the spectral graph wavelet transform and the corresponding
wavelet basis using the spectral decomposition of $\mathcal{L}$ as in \cite{Hammond2009,Tremblay2014}. Let $\Lambda=\left\{  \lambda_{j}\right\}  _{j=1}^{NT}$
be the vector of eigenvalues of the supra-Laplacian
$\mathcal{L}$ satisfying
$\lambda_{1}\leq\lambda_{2}\leq\cdot\cdot\cdot\leq\lambda_{NT}.$
Let $\chi=\left[  \chi_{1}|\chi_{2}|...|\chi_{NT}\right]  $ be the $NT\times
NT$ matrix of column eigenvectors which correspond to
those eigenvalues. 

Denote by $\psi_{s,i}^{t}$ the wavelet at scale $s\in\mathbb{R}^{+}$ centred around node $i\in V$ at time layer $t$. The wavelets are generated by stretching a unique wavelet filter function $g\left(
\boldsymbol{\cdot}\right)  $ by a scale parameters $s>0$ in the network Fourier domain. The matrix representation of the
stretched filter is
\[
\mathcal{G}_{s}=diag\left(  g\left(  s\lambda_{1}\right)  ,g\left(
s\lambda_{2}\right)  ,...,g\left(  s\lambda_{NT}\right)  \right)
\]
that is diagonal on the Fourier modes (the $NT$ eigenvectors of $\mathcal{L}%
$). Hence the wavelet basis at scale $s$ reads as%
\begin{equation}
\Psi_{s}=\left(  \psi_{s,1}^{1}|\psi_{s,2}^{1}|...|\psi_{s,N}^{1}|\psi
_{s,1}^{2}|...|\psi_{s,N}^{T}\right)  =\chi\mathcal{G}_{s}\chi^{\top} ,
\end{equation} where $\psi_{s,i}^{t}$ is the wavelet at scale $s$ centred around node $i$ at the time point $t$.
For a wavelet at scale $s$ centered around node $i$ at time point $t$, we have the relation $\psi_{s,i}^{t}=\chi\mathcal{G}_{s}\chi^{\top}\delta_{i,t}$, which is a column vector of size $NT$. Its value at each node $j$ at time point $p$ is given by $\psi_{s,i}^{t}(j,p)$.
\subsection{Spectral Properties of the Supra-Laplacian Matrix for Community Detection Purposes\label{sec:specL}}
In our context, we interpret each $G^{t}$ as disconnected components and the inter-layer weights as small perturbations. The resulting diagonal blocks of the supra-Laplacian, $\mathcal{L}_{t,t}$, are then perturbed versions of the corresponding Laplacian $L^{t}$. We use Davis-Kahan theorem from matrix Perturbation theory (p.$246$ in~\cite{Stewart1990} and p.$212$ in~\cite{Bhatia1997}) discussed in~\cite{Luxburg} to justify the choice of an eigenvalue around which to center the wavelet filter function. According to the Davis-Kahan theorem, \textit{some} of the first $T$ perturbed eigenvectors of $\mathcal{L}$ are very close to the linear space generated by the vectors $v_{0}^{t}\mathbbm{1}_{G^{t}}$. Here $v_{0}^{t}$ is the eigenvector corresponding to eigenvalue $0$ of matrix $L^{t}$, while $\mathbbm{1}_{G^{t}}$ is the $NT$ zero-padded indicator vector, which has entries $1$ at the node positions of layer $G^t$. 

From spectral graph theory~\cite{Chung1996} it follows that the eigenvector $v_{0}^{t}$ corresponding to the $0$ eigenvalue of the normalized Laplacian matrix $L^{t}$ (of the undirected connected network $G^t$) is not informative of the community structure, since it is equal to the squared node degrees, $D_{t}^{\frac{1}{2}}$. It follows that in the spectrum of the supra-Laplacian there
exists a set of small eigenvalues $\lambda$, whose corresponding eigenvectors are \textit{uninformative} for the community structure within the layers. 

These eigenvalues and their corresponding eigenvectors can only be used to identify each time layer $G^{t}$ as
a separate layer. In fact, the smallest non-zero eigenvalue $\lambda$, 
whose eigenvector is \textit{not spanned} by the set of eigenvectors $v_{0}^{t}$, is sensitive to within-layer connectivity patterns since it may appear as perturbation of the separate layers' Fiedler vectors used for clustering~\cite{Chung1996}. We center our wavelet filter function around this eigenvalue, denoted by $\lambda^{\ast}$, since $\lambda^{\ast}$ is carrier of the coarse description of communities within time layers. We also use $\lambda^{\ast}$ to automatically determine the range of scales $s$, for which relevant communities can be discovered. 

Denote by $\overline{\Lambda}$ the set of smallest eigenvalues whose eigenvectors are well-approximated by the subspace of eigenvectors
$v_{0}^{t}$. According to the Davis-Kahan theorem, the eigenvectors $v$ corresponding to $\lambda\in\overline{\Lambda}$
satisfy
\begin{equation}
\min_{\left\{  \alpha_{t}\right\}  }\left\Vert v-\sum_{t=1}^{T}\alpha_{t}%
v_{0}^{t}\right\Vert \leq\varepsilon\label{vApproximate}%
\end{equation}
for a small $\varepsilon>0$. For the rest of the eigenvalues, the left hand side of this inequality is much larger than $\varepsilon.$ 
Then eigenvalue $\lambda^{\ast}$ is the first eigenvalue which is a perturbation of the Fiedler vectors of the separate time layers, i.e. we have the equality
\begin{equation}
\lambda^{\ast}:=\min\left\{  \lambda:\lambda\in\Lambda\setminus\overline
{\Lambda}\right\}  . \label{lambdaStar}%
\end{equation}

In practice, we discover the position of the
eigenvalue $\lambda^{\ast}$ by solving a series of regression problems: for each of the ordered eigenvectors of the supra-Laplacian $v_{\tau}=\chi_{\tau}$ ($\tau=1,2,...,NT$), we fit the multivariate regression $v_{\tau}=\sum_{t=1}^{T}\beta_{t}\alpha_{t}v_{0}^{t}+\varepsilon_{\tau}$. We select $\lambda^{\ast}$ at the $\tau$ position for which $\left\|\varepsilon_{\tau}\right\|>0.8$, where this bound was empirically selected. Since $\lambda^{\ast}\leq\lambda_{T+1}$, we have to solve
at most $T+1$ regression problems in order to find the position of $\lambda^{\ast}.$

\subsection{Graph Wavelet Filter $g$ via
$B$-Splines and Parameter Selection \label{sec:gBsplines}}

We propose a new wavelet filter function $g$ modeled as a cubic $B-$spline
\cite{DeBoor2001} with appropriately chosen knots. This function is not only smooth but also has a compact support. Namely, we put
\begin{equation}
g\left(  y\right)  :=B_{3}\left(  0,y_{1},y_{2}%
,y_{3},y_{4};y\right)  \label{gJK}%
\end{equation}
where for the knots of the cubic $B-$spline $B_{3}$ we have
\begin{equation}
0<y_{1}<y_{2}=y_{3}<y_{4}
\end{equation}
and the function $g$ is zero out of the interval $\left[  0,y_{4}\right]$. By the properties of $B-$splines, $B_{3}>0$ for $y\in\left(  0,y_{4}\right)
.$ As indicated, this spline function has a double knot at $y_{2}=y_{3}.$ Function $g$ inherits the basic properties of $B$-splines \cite{DeBoor2001}, including good 
properties of the Fourier coefficients of $g$ since $g(y)$ may be extended for $y<0$ and $y>y_{4}$ periodically 
to belong to $C^{1}$, which is important for the invertibility of the Fourier wavelet transforms. Other functions can further be pursued depending on the application at hand, and possible options have been reviewed in~\cite{Leonardi2013}. 

In the following we describe how to choose parameters $y_{1},$ $y_{2}=y_{3},$
and $y_{4}$ of the wavelet filter function $g,$ and the range of scales $s$
relevant for multi-scale communities within and across layers of
the temporal network. Some of the arguments we make are the same as in \cite{Hammond2009,Tremblay2014}.
However, we adapt these to the nature of $g$ and the aim of centering it around an appropriate eigenvalue.

First, the maximum scale
$s_{\max}$ is set so that the filter function $g\left(  s_{\max}y\right)  $
starts decaying as a power law only after $y=\lambda^{\ast}$, hence $\lambda^{\ast}s_{\max}=y_{2}=y_{3}.$
Second, we need to keep a part of the corresponding eigenvector
$\chi_{\lambda^{\ast}}$ in the wavelets of every scale, so that all wavelets
are sensitive to the large scale community structure within each time point.
We propose as minimum scale $s_{\min}$ the one for which $g\left(  s_{\min
}\lambda^{\ast}\right)  $ becomes smaller than $1$. Therefore, $s_{\min}\lambda^{\ast}=y_{1}.$
We also impose that $g\left(  s_{\min}\boldsymbol{\cdot}\right)  $ spans at
least the whole range of eigenvalues between $0$ and $1$ which implies $s_{\min}\times1=y_{2}.$

We require that the filter at the maximum scale be highly
selective around $\lambda^{\ast}$. For this purpose all other eigenvalues and especially
$\lambda^{q+1}$, where we have put $\lambda^{\ast}=\lambda^{q}$, have to be attenuated. Choosing an attenuation by a factor
$10$, leads to $g\left(  s_{\max}\lambda^{q}\right)  =10\left(  s_{\max}\lambda^{q+1}\right)$. We thereby ensure that the filter at the maximum scale
essentially keeps the information from $\chi_{\lambda^{\ast}}$.

This argumentation gives us spectrum adapted equations for $s_{\min}$,
$s_{\max}:$%

\begin{equation}
s_{\min}=\frac{y_{1}}{\lambda^{\ast}},\quad y_{2}=y_{3}=\frac{y_{1}}%
{\lambda^{\ast}},\quad s_{\max}=\frac{y_{1}}{\left(  \lambda^{\ast}\right)
^{2}} ,
\end{equation}
where we see that $y_{1}$ has the unique effect of translating the scale
boundaries $s_{\min}$ and $s_{\max}$ on the $%
\mathbb{R}
^{+}$ axis. Therefore, $y_{1}$ can be safely fixed to $1$, i.e. $y_{1}=1$. \sloppy Finally, similar to the approach in \cite{Hammond2009,Tremblay2014} we choose a
logarithmically spaced sampling of $M$ scales between the scale boundaries
$s_{\min}$ and $s_{\max}$: $S=\left\{  s_{1}=s_{\min,}s_{2},...,s_{M}=s_{\max
}\right\}  $. 

\begin{figure}[tb]
\subfigure[$g$ centred around $\lambda_{2}$]{
\includegraphics[width=.48\columnwidth]{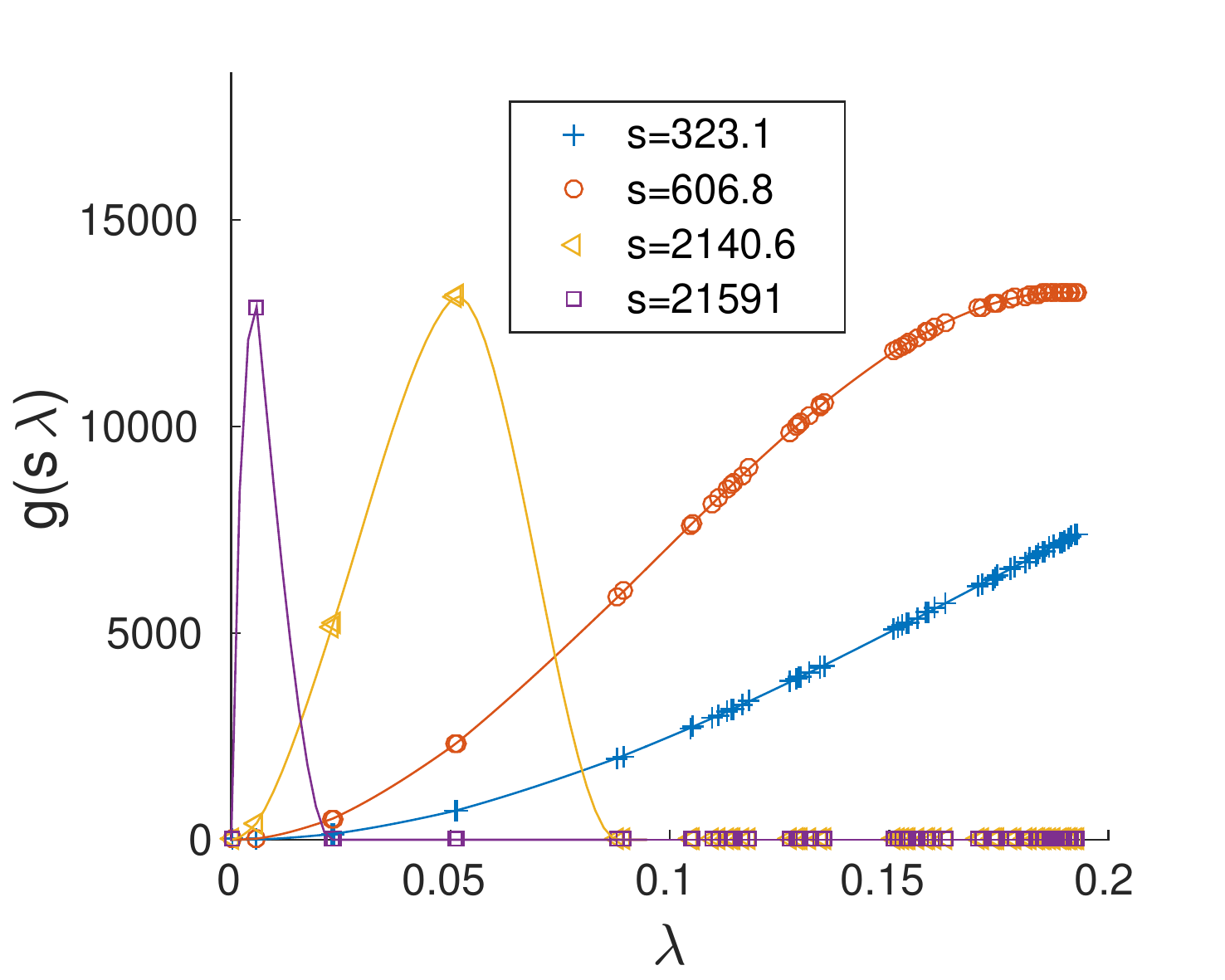}\label{fig:filter2:a}}
\subfigure[$g$ centred around $\lambda^{\ast}$]{
\includegraphics[width=.48\columnwidth]{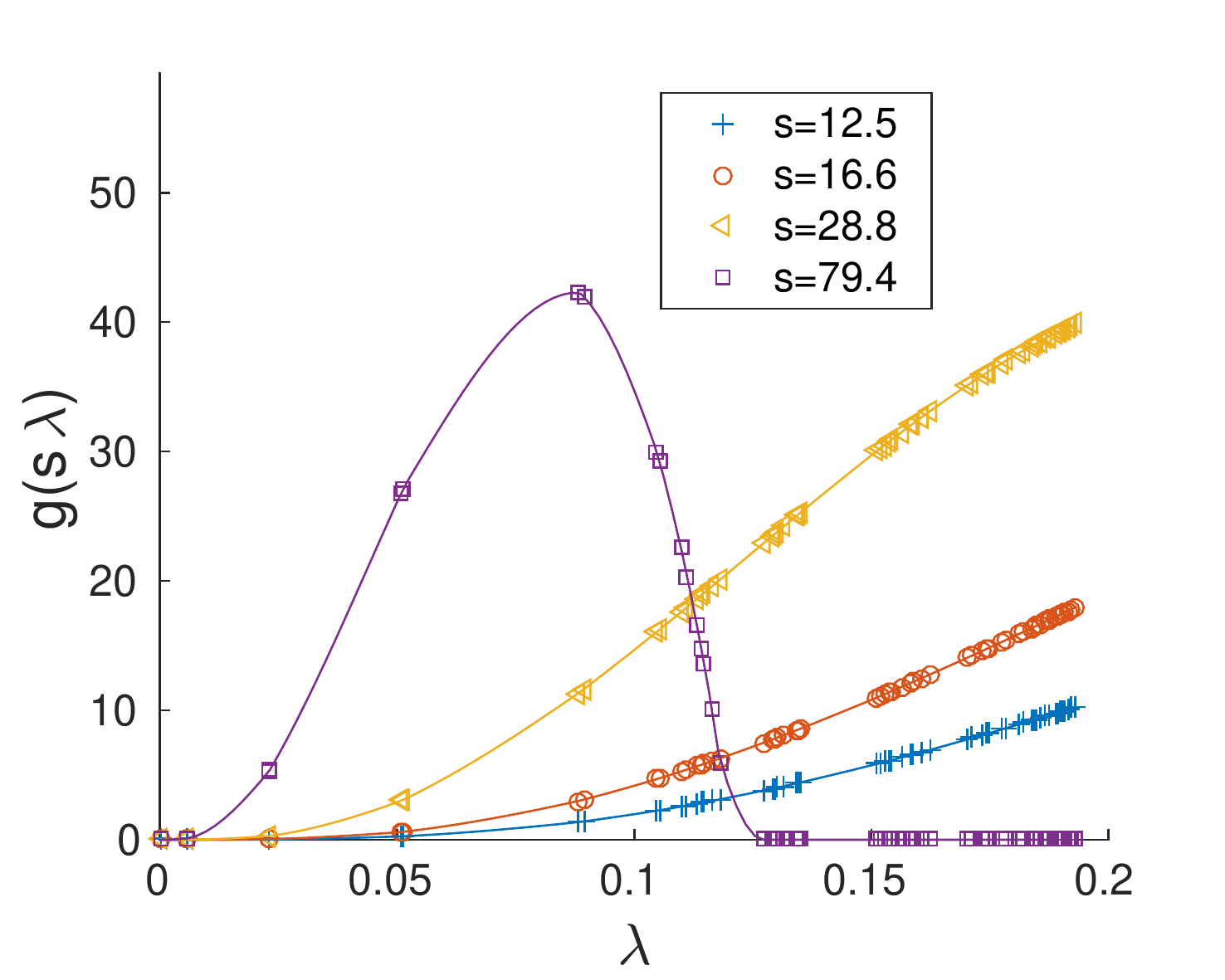}\label{fig:filter2:b}}
\caption[$g$ filter based on a $B$-spline for a temporal network.]{ 
$B$-spline based filter function $g$ for four different scales $s$.
The actual eigenvalues $\lambda$ are obtained from a temporal multi-scale benchmark whose communities at large scales change over time (discussed in Section~\ref{sec:experiemnts}). The temporal network has $N=640$ nodes for each of the $T=33$ time
layers with Sales-Pardo parameters $\rho=1$ and $\bar{k}=16$. In (a) filter $g$ is centred around $\lambda_{2}$ as originally
proposed in \cite{Tremblay2014} for monoplex networks. In (b)
the filter $g$ is centred around $\lambda^{\ast}$, which is $\lambda_{10}$ obtained as proposed in Section~\ref{sec:specL}. In total $50$ scales were obtained
and the four visualized scales correspond to the $7^{th},13^{th},25^{th}$ and
$47^{th}$ scale. }%
\label{fig:filter2}%
\end{figure}

Rather than fixing the first
parameter $y_{2}$ of filter $g$ around the second eigenvalue $\lambda_{2}$ of
the supra-Laplacian, we fix $y_{2}$ centred around $\lambda^{\ast}$. Thus we
\textit{attenuate} the role of the eigenvalues $\lambda_{j}\in\Lambda$ with
$\lambda_{j}<\lambda^{\ast},$ since as we explained above the eigenvectors of these
eigenvalues $\lambda_{j}$ are not relevant for discovering community
structures prevalent at each time point.

In Figure \ref{fig:filter2} we visualize function $g$ for a \textit{temporal network} and compare the shape of $g$ when
it is centred around $\lambda_{2}$ and around $\lambda^{\ast}$ -- obtained as proposed. The eigenvalues were obtained from a multi-scale benchmark temporal network whose communities at large scales change over time.

\subsection{Agglomerative Connectivity-Constrained Clustering and Detection of
Stable Partitions\label{sec:stability}}

For small scales $s$, $\psi_{s,i}^{t}$ is localized around the direct neighbours of $i$ in layer $t$ and to few nodes in
neighbouring time layers. With an increasing scale $s$, $\psi_{s,i}^{t}$
spreads to a larger neighbourhood which eventually becomes the whole multilayer
network. Hence, we use $\psi_{s,i}^{t}$ as a feature vector for $i_{t}$
at scale $s$. 

Similar to \cite{Tremblay2014}, we determine the distance $D_{s}(i_{t},j_{p})$ between nodes $i_{t}$ and $j_{p}$ ($i,j=1,2,3,...,N$;
$t,p=1,2,3,...,T$) at scale $s$ using the correlation distance between wavelets $\psi_{s,i}^{t}$ and $\psi_{s,j}^{p}$. We speed up computations of the full spectrum of $\mathcal{L}$ and all $D_{s}(i_{t},j_{p})$ using approximations proposed in \cite{Hammond2009} and \cite{Tremblay2014}. 

We cluster nodes into communities using distances $D_{s}(i_{t},j_{p})$ and an agglomerative connectivity-constrained clustering procedure \cite{Pons,Kuncheva2015} with ``average'' linkage. In this way we respect the time-ordered structure of the temporal network  since nodes in the same time layer or nodes across neighbouring time layers are considered first for merging. We obtain the partition at scale $s$, $P_{s}$, by cutting the resulting dendrogram at a height equal to the average of the maximal gaps of all the root-leaf paths of the dendrogram \cite{Tremblay2014}. Repeating the above for all $s\in S$, we obtain the multi-scale set of partitions $\mathcal{P}=\left\{  P_{s}\right\}_{s\in S}$. We calculate the stability $\gamma_{a}(s)$ of the partition at a given scale $s$ using the approach outlined in \cite{Tremblay2014}.
\section{Experimental Results\label{sec:experiemnts}}

In this section we provide simulation experiments to measure 
the performance of the TMSCD method on two types of benchmarks for temporal networks in comparison to the performance of the
modularity maximization (MM) method \cite{Mucha}, for different resolution parameter values $\gamma$ on the same set of benchmarks. 

The first type of benchmark networks we use is a further contribution of the present work, since we identify three classes of temporal networks which may serve as \textit{benchmarks}
for multi-scale community detection on temporal networks. We construct these benchmarks as time-varying Sales-Pardo (SP) networks~\cite{Sales-Pardo2007}. An SP has three scales of communities based on which the network is constructed using parameters $\rho$ (quantifies how separated the three scales are) and $\overline{k}$ (the average node degree that controls how dense the network is).
For a given length of the temporal network $T$, we generate SP multi-scale community structures that merge and split over time. Based on these community structures, we simulate a time-ordered sequence of Sales-Pardo networks. The three classes are determined by the scale at which the change occurs: small scale (SSC), medium scale (MSC) or large scale (LSC) change over time. The second type of benchmarks, proposed in~\cite{Granell2015}, have one ``true'' partition at each time point.

The performance of a given 
algorithm at a given scale (resolution) is measured as the maximum value of the adjusted rand index (ARI) \cite{Hubert1985}
between the ``true'' partition of the benchmark and the partition $P$ at this scale (resolution). Since the SP benchmarks have three true partitions
corresponding to three different scales, we refer to the large (resp.
medium, small) scale as LS (resp. MS, SS). We also investigate the
performance of TMSCD and MM on benchmarks produced using different
values of $\rho$ and $\overline{k}$. 

For the TMSCD method, the instability $1-\gamma_{a}$ at scale $s$ is obtained as outlined in \cite{Tremblay2014}. The smaller $1-\gamma_{a}$, the more stable is the community partition for scale $s$. For the MM method, the instability for a
resolution parameter $\gamma$ is obtained as described in \cite{Delvenne2010}
and is measured by the normalized variation of information (VI) metric \cite{Meila2007}. The smaller VI,
the more stable is the community partition at resolution parameter $\gamma$.

\subsection{Comparative Results on Temporal Benchmark Networks with Multi-Scale
Community Structure\label{sec:MSresults}}
\subsubsection{Discussion on effect of inter-layer weights on the performance of TMSCD and MM.} 

First, we illustrate the performance of the TMSCD and MM method on an SSC, MSC and LSC network with $T=21$, $T=17$ and $T=33$ time layers, where $\rho=1$, $\bar{k}=16$, and $N=640$ at each time point. For both methods, we use different fixed inter-layer weights $\omega=0.5,1,2,5,10$ and
the LART-type inter-layer weight $\omega_{i}^{t,t+1}$ proposed in Section
\ref{sec:weight}, which we refer to as $\omega=LART$.

For TMSCD, we set $M=50$ scales $s\in S$, and for MM we manually set $60$ values of resolution parameters $\gamma$ in the
interval $[0.05,40]$ such that there are more values in the interval $\left[
0.05,1\right]$. For both methods we use $20$ repetitions to obtain
instability $\gamma_{a}\left(  s\right)$ at scale $s$ and $VI(\gamma)$ at each resolution $\gamma$.

\begin{figure}
\subfigure{
\includegraphics[width=\columnwidth]{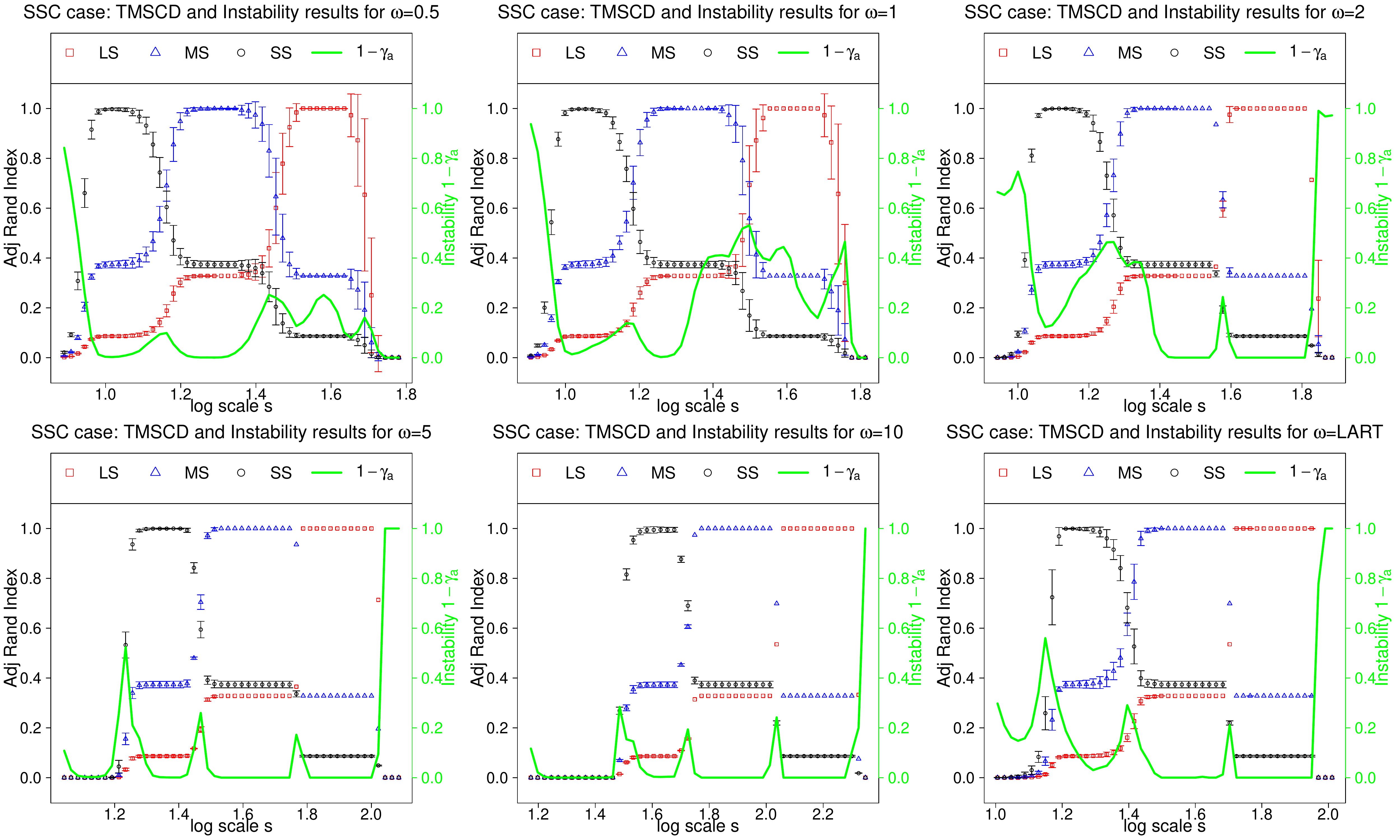}}\\
\subfigure{
\includegraphics[width=\columnwidth]{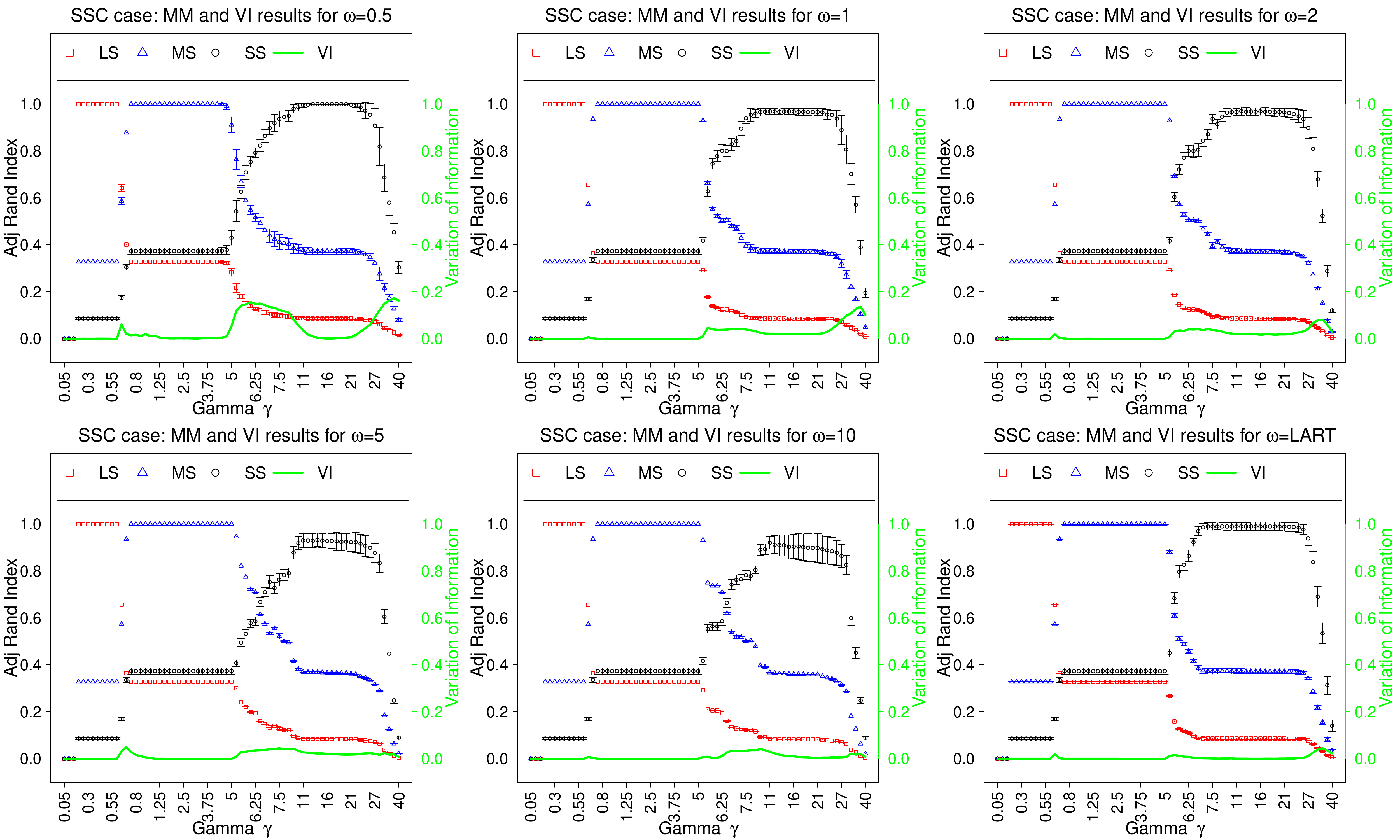}}
\caption[TMSCD and MM results and instability on SSC benchmark networks.]%
{\textit{TMSCD (top) and MM (bottom) results for SSC multi-scale benchmark network. } Each pair of plots corresponds to different inter-layer weights $\omega$. First, we plot the results of TMSCD and MM on a realization of SSC. Each scale
outputs a partition for all nodes across all time points. For each scale $s$,
we plot the similarity with the small (SS) (medium (MS), large (LS))
theoretical scale, computed as the average over all time
points including std.dev. error bars. We observe scales where the
exact small (resp. medium, large) scale theoretical partition is uncovered.
Second, for TMSCD we plot instability $1-\gamma_{a}$ versus scale $s$; for MM we
plot variation of information (VI) versus parameter $\gamma$. The
associated partitions of scales with low instability corresponding to the theoretical partitions. }%
\label{fig:SSC_Results}%
\end{figure}

\begin{figure}
\subfigure{
\includegraphics[width=\columnwidth]{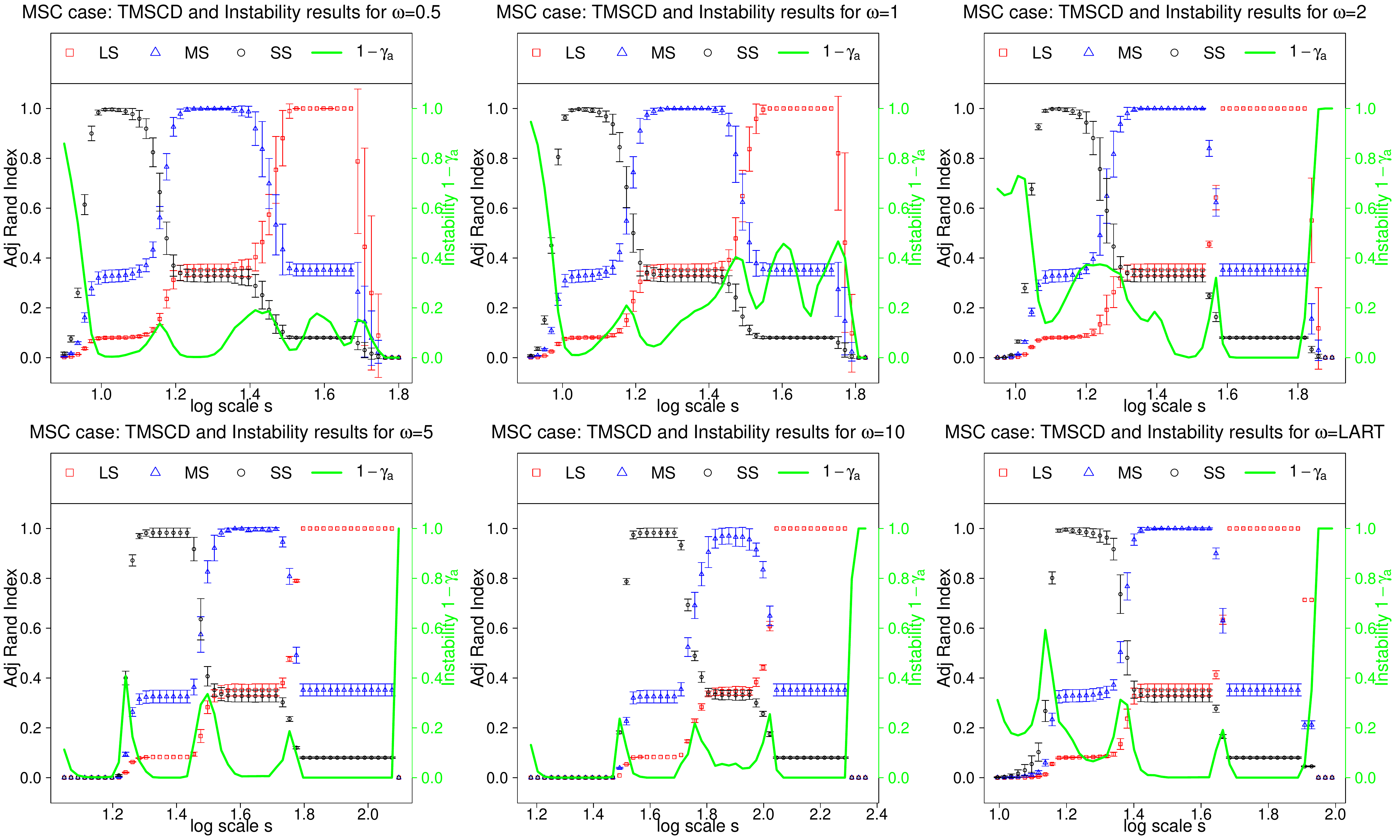}}\\
\subfigure{
\includegraphics[width=\columnwidth]{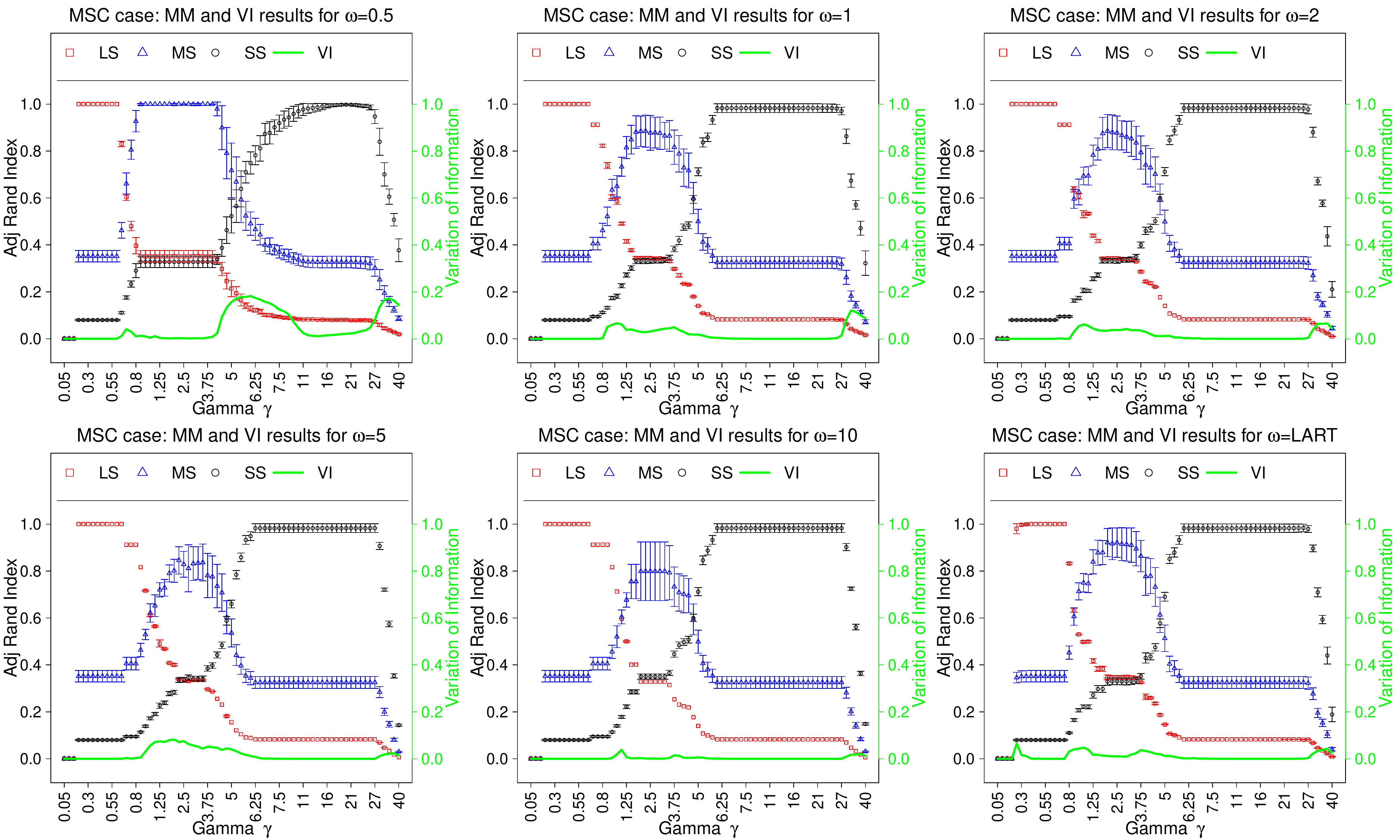}}
\caption[TMSCD and MM results and instability on MSC benchmark networks.]%
{\textit{TMSCD (top) and MM (bottom) results for MSC multi-scale benchmark network. } Each pair of plots corresponds to different inter-layer weights $\omega$. First, we plot the results of TMSCD and MM on a realization of MSC. Each scale
outputs a partition for all nodes across all time points. For each scale $s$,
we plot the similarity with the small (SS) (medium (MS), large (LS))
theoretical scale, computed as the average over all time
points including std.dev. error bars. We observe scales where the
exact small (resp. medium, large) scale theoretical partition is uncovered.
Second, for TMSCD we plot instability $1-\gamma_{a}$ versus scale $s$; for MM we
plot variation of information (VI) versus parameter $\gamma$. The
associated partitions of scales with low instability corresponding to the theoretical partitions. }%
\label{fig:MSC_Results}%
\end{figure}

\begin{figure}
\subfigure{
\includegraphics[width=\columnwidth]{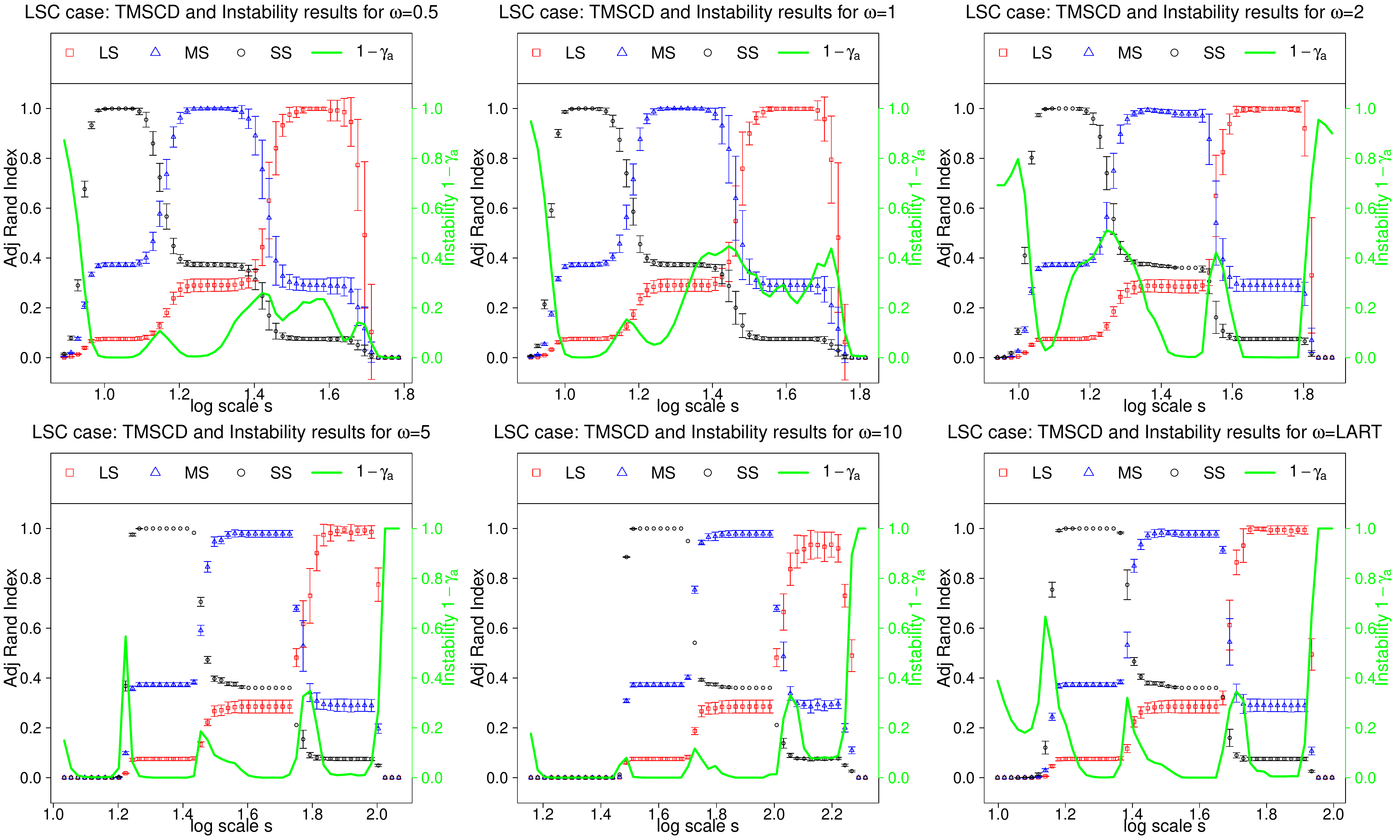}}\\
\subfigure{
\includegraphics[width=\columnwidth]{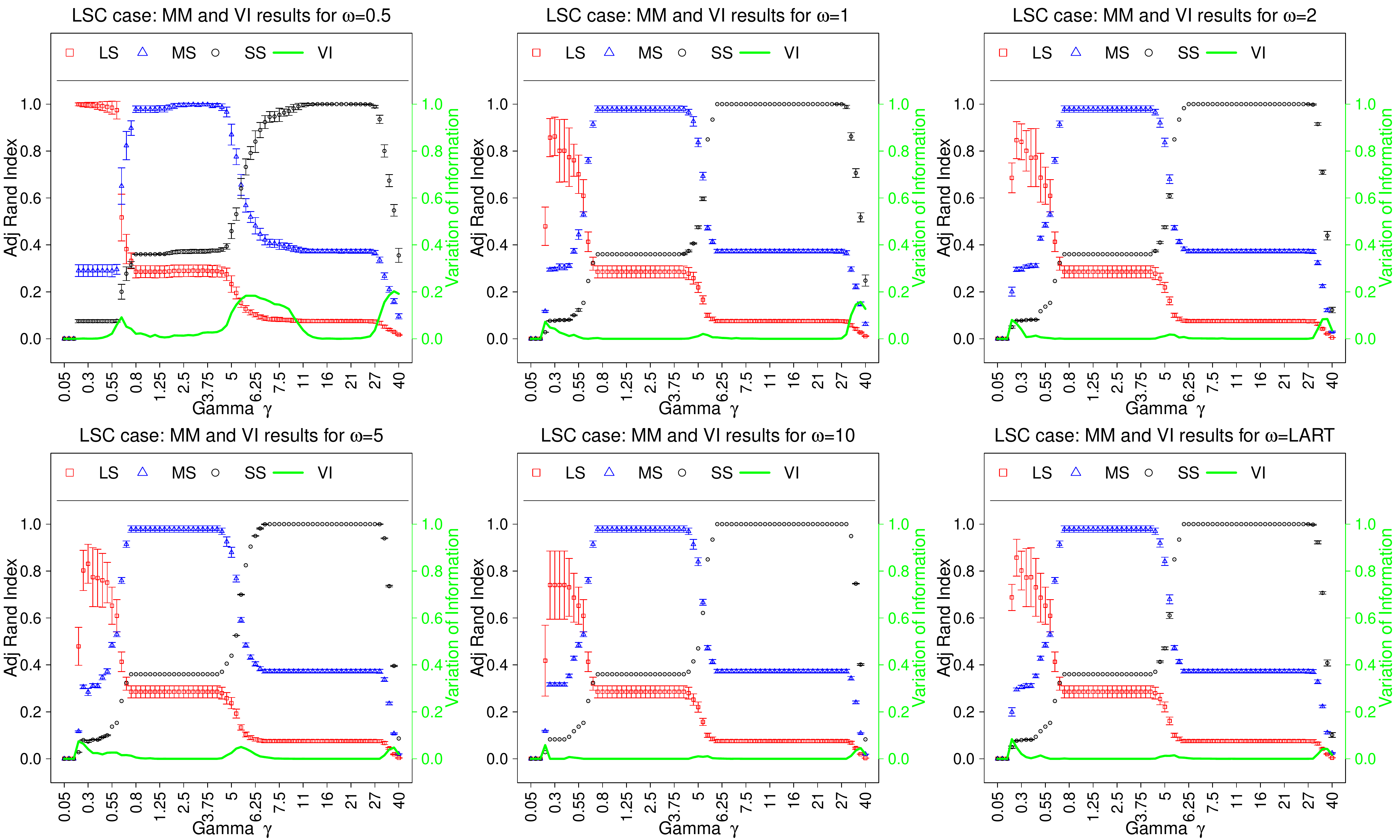}}
\caption[TMSCD and MM results and instability on LSC benchmark networks.]%
{\textit{TMSCD (top) and MM (bottom) results for LSC multi-scale benchmark network. } Each pair of plots corresponds to different inter-layer weights $\omega$. First, we plot the results of TMSCD and MM on a realization of LSC. Each scale
outputs a partition for all nodes across all time points. For each scale $s$,
we plot the similarity with the small (SS) (medium (MS), large (LS))
theoretical scale, computed as the average over all time
points including std.dev. error bars. We observe scales where the
exact small (resp. medium, large) scale theoretical partition is uncovered.
Second, for TMSCD we plot instability $1-\gamma_{a}$ versus scale $s$; for MM we
plot variation of information (VI) versus parameter $\gamma$. The
associated partitions of scales with low instability corresponding to the theoretical partitions. }%
\label{fig:LSC_Results}%
\end{figure}

The results of TMSCD and MM on a realization of the SSC (resp. MSC, LSC) and the instabilities
$1-\gamma_{a}$ versus scale $s$ (VI versus parameter $\gamma$) for different weights
$\omega$ are presented in Figure \ref{fig:SSC_Results} (resp. Figure \ref{fig:MSC_Results} and Figure \ref{fig:LSC_Results}). 

Overall, the TMSCD recovers perfectly communities at all three
scales and inter-layer weights $\omega$ have almost no effect over the
results. For small $\omega$ ($\omega=0.5,1,2$) instability is high, but for $\omega=LART$ and $\omega=5,10$ the associated partitions of
scales with low instability correspond to the true
partitions. For large $\omega$ ($\omega=5,10$ and $\omega=LART$) a fourth
stable scale appears at the smallest $s$. This is
stable for $\omega=5,10$ but unstable for $\omega=LART$, which signifies the importance of carefully selected weights. MM recovers perfectly
communities at LS and MS, but there is increased variability at recovering
communities at SS for an increasing inter-layer weight $\omega$ ($\omega
=5,10$), The instability of MM is not as
sensitive as the one that is used for TMSCD.

To conclude, using
$\omega=LART$ inter-layer weights appears to provide us with low instability
only at the true partitions. This includes a higher instability at the fourth
scale which appears for larger $\omega$ ($\omega=10$) - the partition at this scale is formed for $N$
communities, and each community is formed by the set of nodes $\left\{
i_{t}:t=1,2,...,T\right\}$. As discussed in Section~\ref{sec:weight}, this phenomenon is a results of larger
inter-layer weights, which affect more the properties of a node $i$ at layer $t$ than its
within-layer connections which have an average node degree $\bar{k}=16$. In contrast, the LART weights, in the range $[0,8]$, follow a bell-shaped distribution centred around $4$. Thus they do not interfere with the within-layer connections and support the community detection process over time. 

On the other hand, the instability procedure for MM
is not as sensitive. In the case of real data it would be challenging to select parameters $\gamma$ for which to investigate community
partitions. 

\begin{figure}\centering
\subfigure[Results for SSC temporal benchmark]{
\includegraphics[height=0.25\textheight,width=0.7\textwidth]{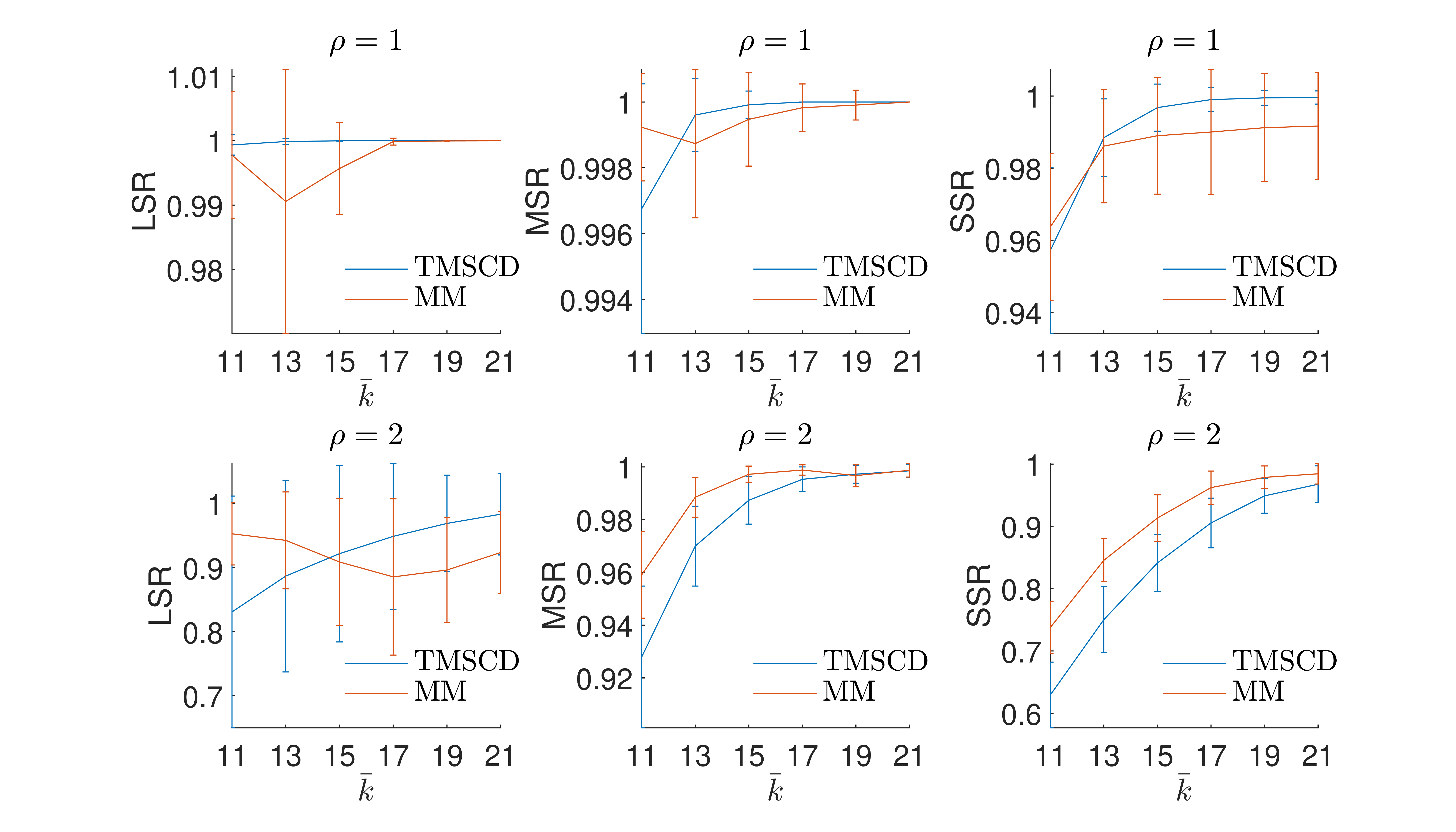}\label{fig:LSC_rho_Results:a}}
\subfigure[Results for MSC temporal benchmark]{
\includegraphics[height=0.25\textheight,width=0.7\textwidth]{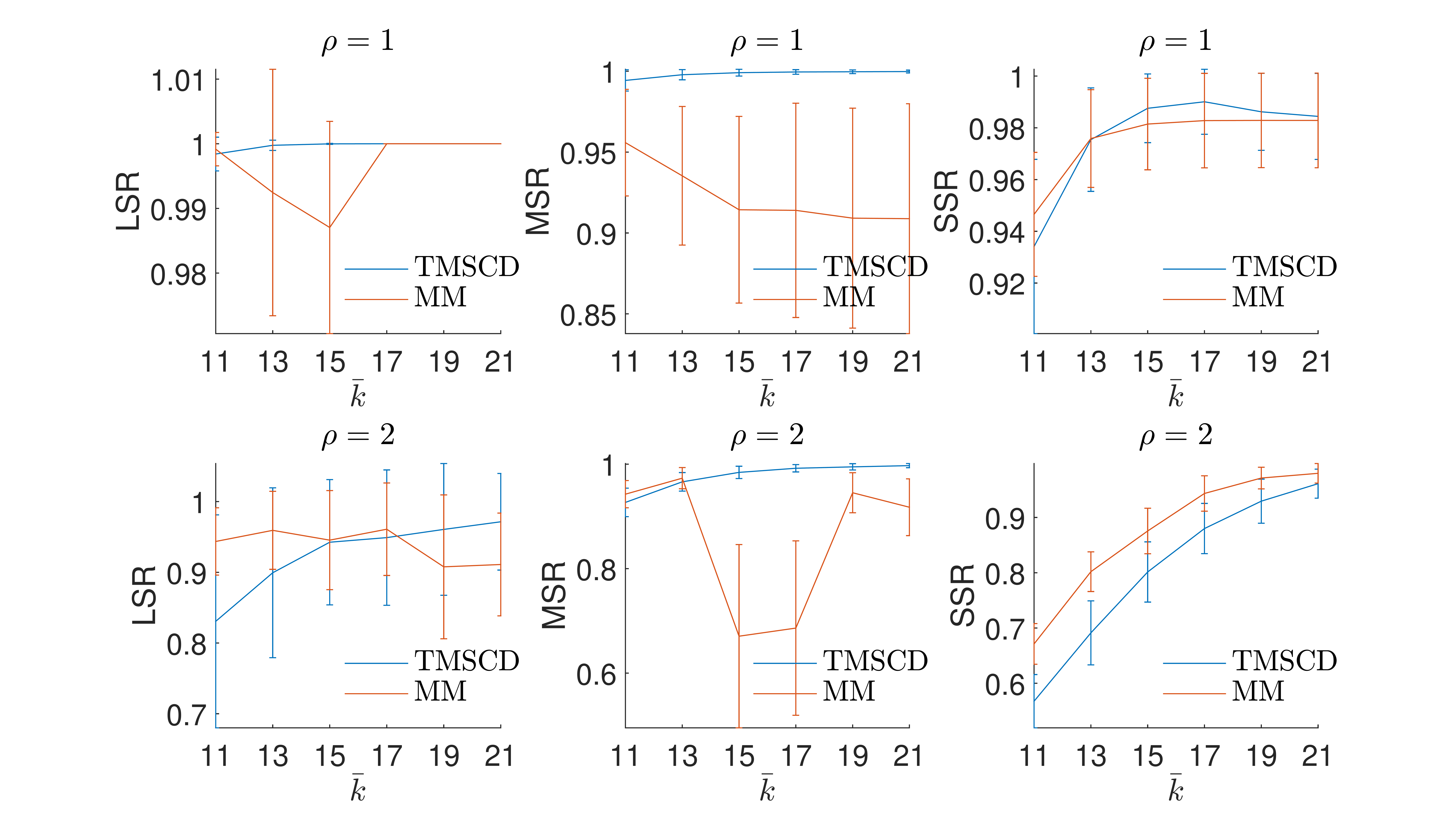}\label{fig:LSC_rho_Results:b}}
\subfigure[Results for LSC temporal benchmark]{
\includegraphics[height=0.25\textheight,width=0.7\textwidth]{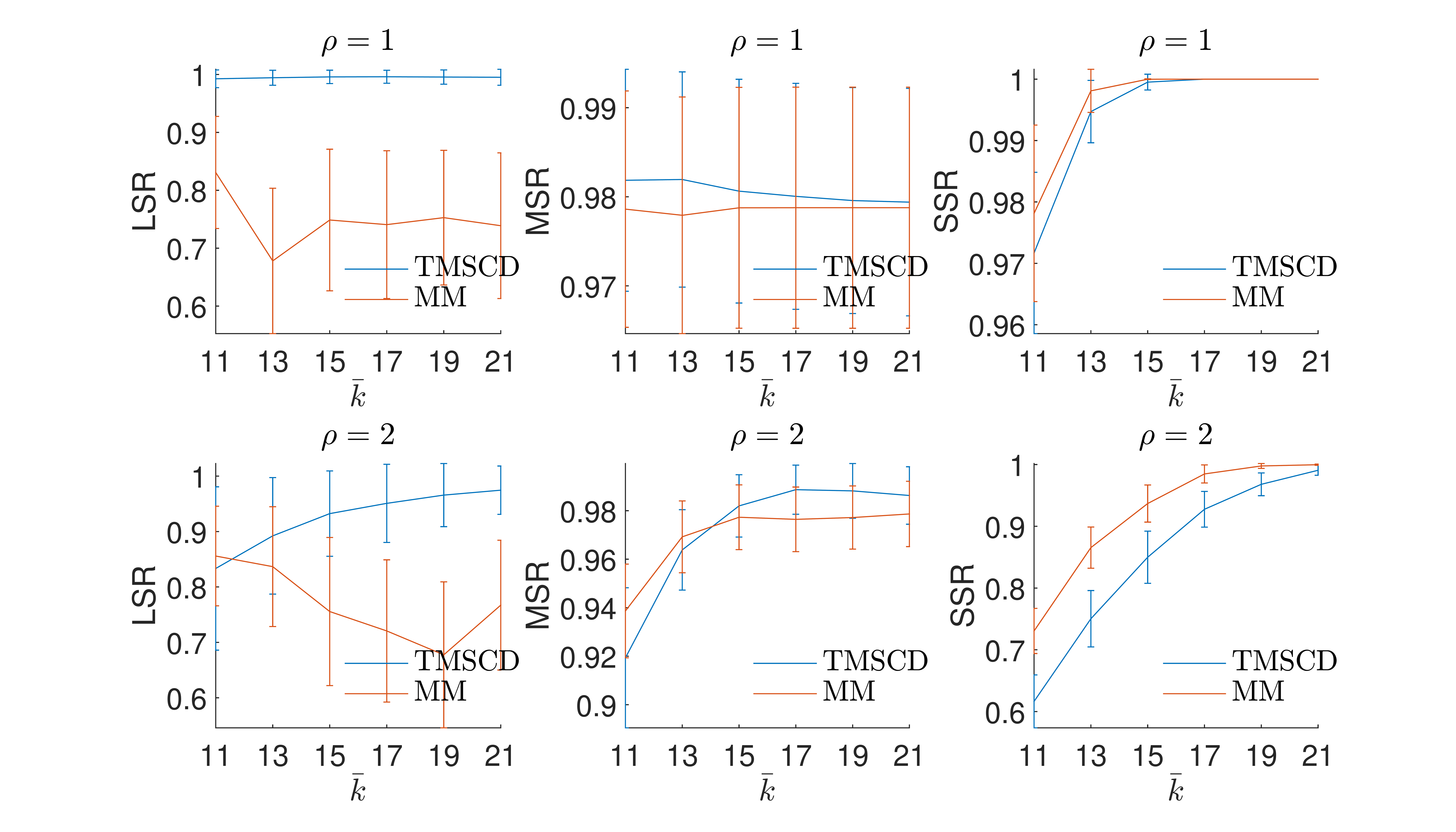}\label{fig:LSC_rho_Results:c}}
\caption{Comparison between the LSR, MSR and SSR values obtained for the TMSCD and MM
multi-scale community mining methods on (a) SSC, (b) MSC and (c) LSC temporal benchmark network for
different parameters: left (resp. right) column for $\rho=1$ (resp. $\rho=2$)
and different values of $\bar{k}$. We plot the average and the $\pm$ one standard deviation
over the five best results for each of the $50$ realizations of each type of
network for each set of parameters.}%
\label{fig:LSC_rho_Results}%
\end{figure}
\subsubsection{Discussion on overall performance of TMSCD and MM.} 

We compare TMSCD and MM for different sets of
parameters $\rho=1,2$ and $\bar{k}=11,13,15,17,19,21$, where we set LART-type inter-layer weights. We compare the
obtained communities to the ground truth for $50$ realizations of the SSC
(Figure~\ref{fig:LSC_rho_Results:a}), MSC (Figure~\ref{fig:LSC_rho_Results:b}) and
LSC (Figure~\ref{fig:LSC_rho_Results:c}). For each combination $(\rho,\bar{k})$, the large scale rate (LSR) (resp. medium (MSR) and small (SSR) scale rates) indicates the success rate of the communities found by TMSCD and MM being compared to the large (resp. medium, small) scale ground truth community structure. The success rate is the average over the top five adjusted rand index (ARI) values over all scales $s$ or parameter values $\gamma$.

For $\rho=1$, both methods perform equally well in all three cases with almost full recovery of communities at all scales. In some cases, TMSCD has slight advantage of recovering MS and LS communities. 
For $\rho=2$, the performance of both methods decreases for small $\bar{k}$. Both methods perform equally well at recovering MS communities, but we can note that TMSCD performs better
at recovering LS communities for larger $\bar{k}$.  

Overall TMSCD performs slightly better than MM and has much smaller deviation
in the final results. In general, uncovering communities
when $\rho=2$ is harder since $\rho$ controls how separated are the
communities at the three scales. When $\rho$ is larger, the separation of the
communities is not as clear, so SS communities cannot be distinguished easily.
Furthermore, we note that when $\bar{k}$ is small nodes have fewer edges and
it is difficult for both methods to uncover SS communities since they fade in
the MS communities.

\subsection{Comparative Results on Benchmarks with One True
Partition\label{sec:Arenasresults}}

We compare the performance of TMSCD and MM on the \textit{Grow, Merge
and Mixed} benchmark networks proposed in \cite{Granell2015}, with default model parameters, and we set $T=100$ and $N=128$. 

First, we illustrate the performance of the TMSCD and MM method on an \textit{Grow, Merge
and Mixed} network. For both methods, we use different fixed inter-layer weights $\omega=0.5,1,2,5,10$ and
the LART-type inter-layer weight $\omega_{i}^{t,t+1}$ proposed in Section
\ref{sec:weight}.

\begin{figure}
\subfigure{
\includegraphics[width=\columnwidth]{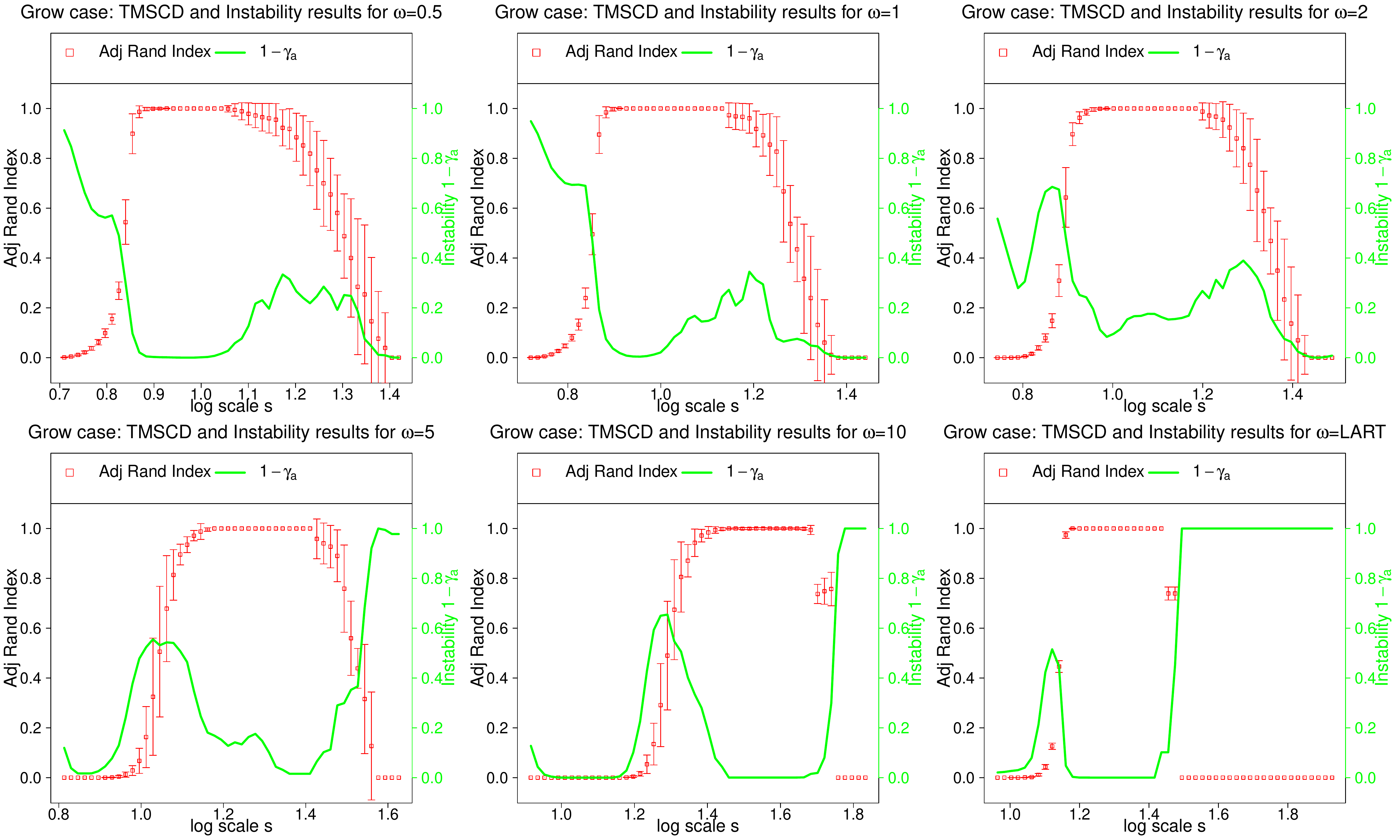}}\\
\subfigure{
\includegraphics[width=\columnwidth]{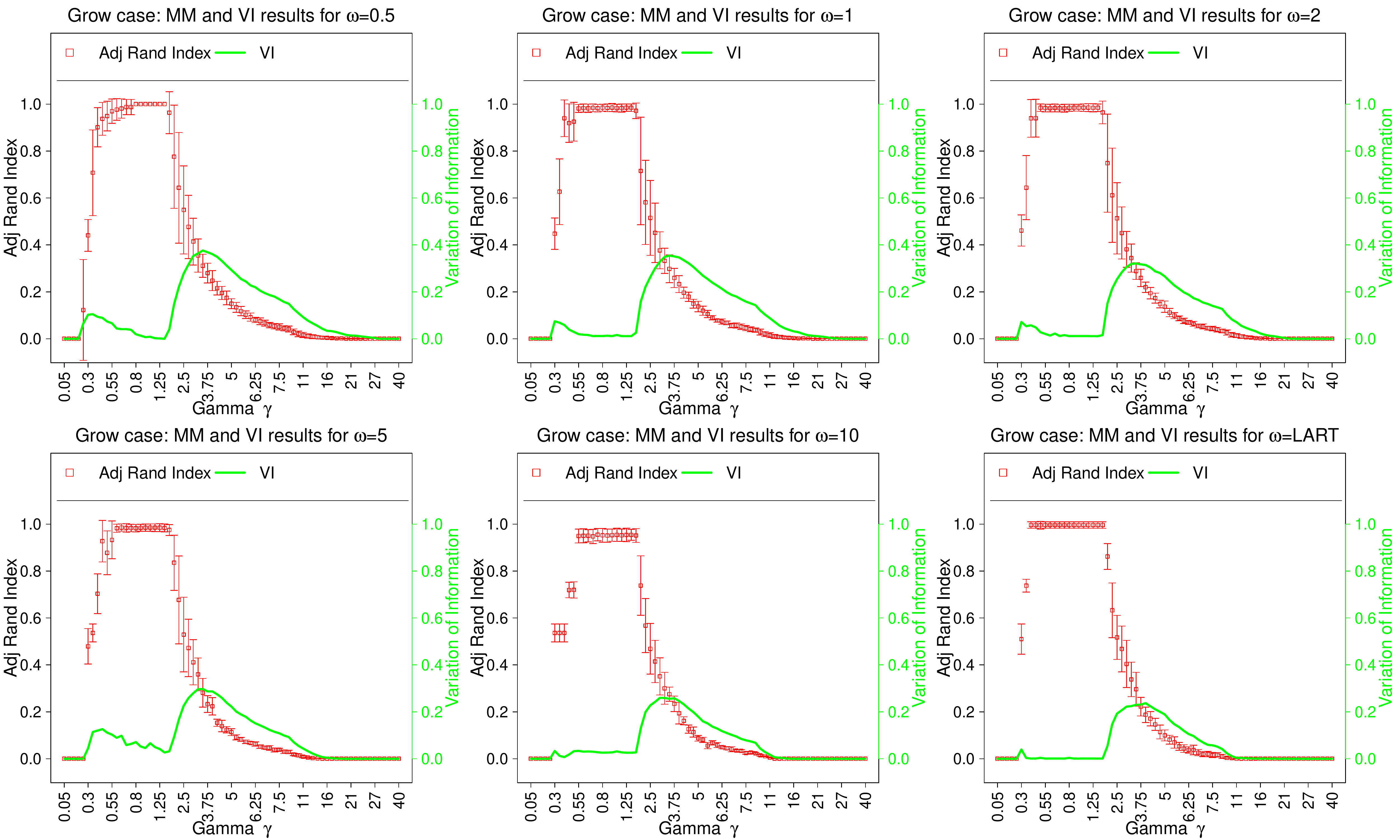}}
\caption[TMSCD and MM results and instability on Grow benchmark networks.]%
{\textit{TMSCD (top) and MM (bottom) results for Grow multi-scale benchmark network.} Each pair of plots corresponds to different inter-layer weights $\omega$. First, we plot the results of TMSCD and MM on a realization of Grow model. Each scale
outputs a partition for all nodes across all time points. For each scale $s$,
we plot the similarity with the theoretical scale, computed as the average over all time
points including std.dev. error bars. In this way, we observe scales where the
exact small (resp. medium, large) scale theoretical partition is uncovered.
Second, for TMSCD we plot the instabilities $1-\gamma_{a}$ versus scale $s$; for MM we
plot variation of information (VI) for each resolution parameter $\gamma$. The
associated partitions of scales with low instability (i.e. high stability)
corresponding to the theoretical partitions. }%
\label{fig:Grow_Results}%
\end{figure}

\begin{figure}
\subfigure{
\includegraphics[width=\columnwidth]{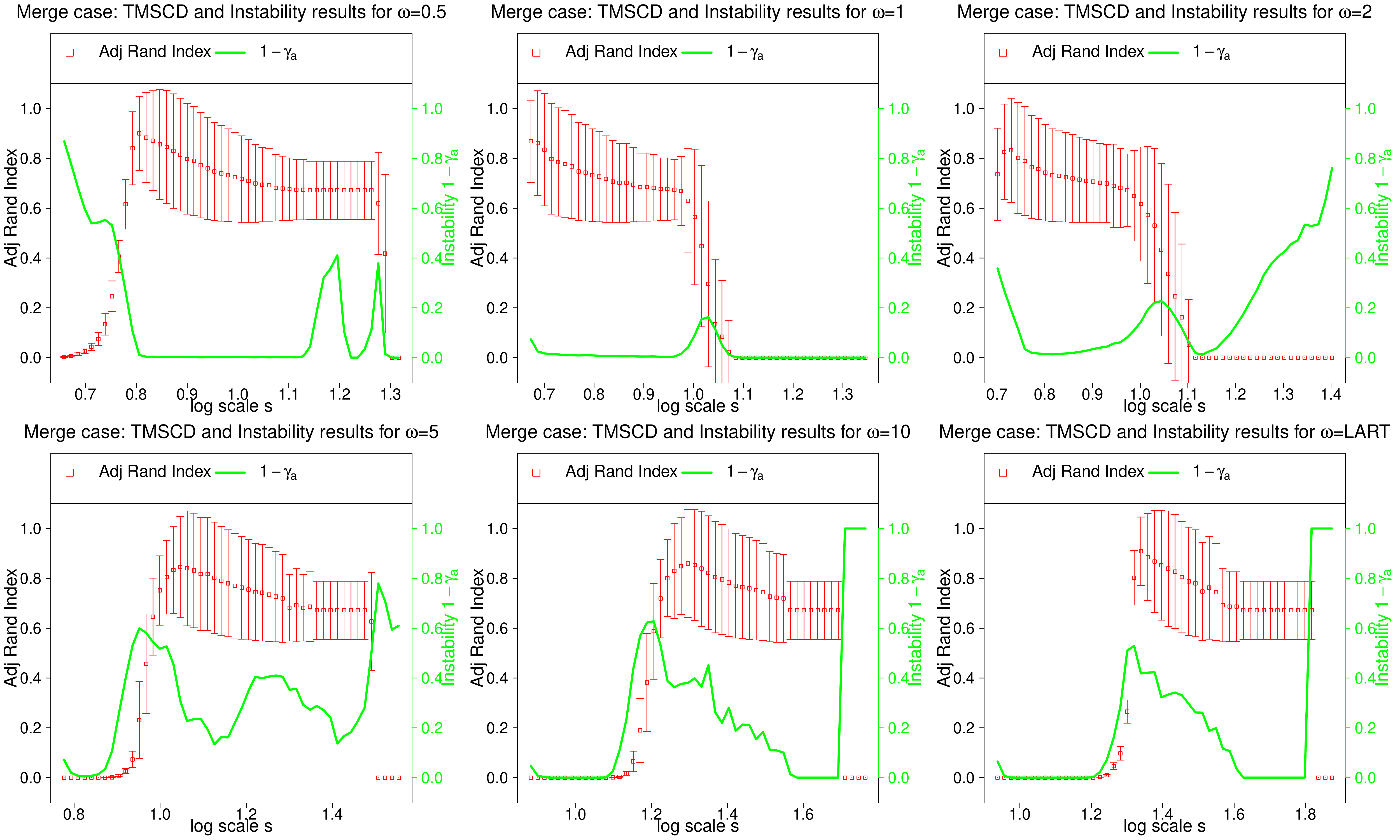}}\\
\subfigure{
\includegraphics[width=\columnwidth]{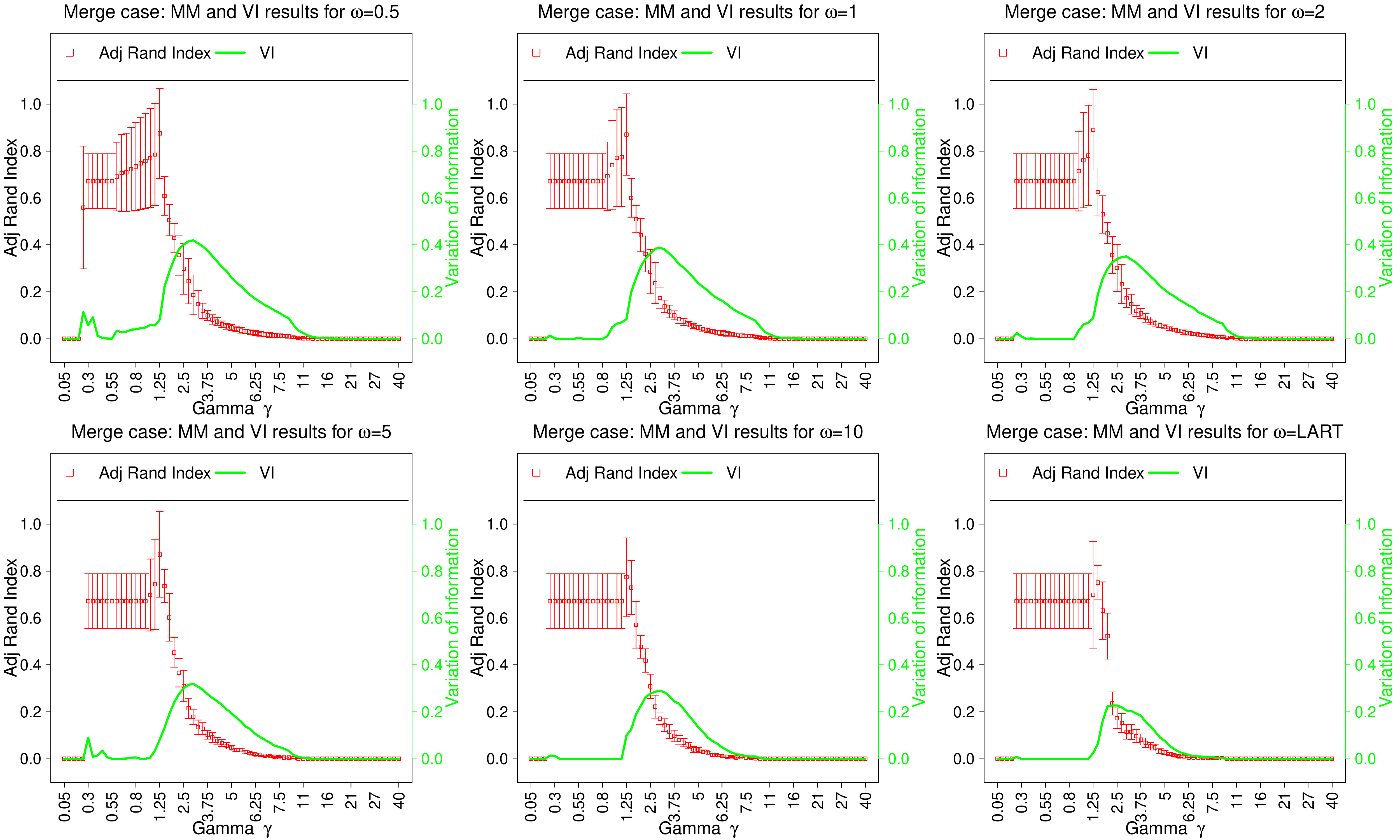}}
\caption[TMSCD and MM results and instability on Merge benchmark networks.]%
{\textit{TMSCD (top) and MM (bottom) results for Merge multi-scale benchmark network.} Each pair of plots corresponds to different inter-layer weights $\omega$. First, we plot the results of TMSCD and MM on a realization of Merge model. Each scale
outputs a partition for all nodes across all time points. For each scale $s$,
we plot the similarity with the theoretical scale, computed as the average over all time
points including std.dev. error bars. In this way, we observe scales where the
exact small (resp. medium, large) scale theoretical partition is uncovered.
Second, for TMSCD we plot the instabilities $1-\gamma_{a}$ versus scale $s$; for MM we
plot variation of information (VI) for each resolution parameter $\gamma$. The
associated partitions of scales with low instability (i.e. high stability)
corresponding to the theoretical partitions. }%
\label{fig:Merge_Results}%
\end{figure}

\begin{figure}
\subfigure{
\includegraphics[width=\columnwidth]{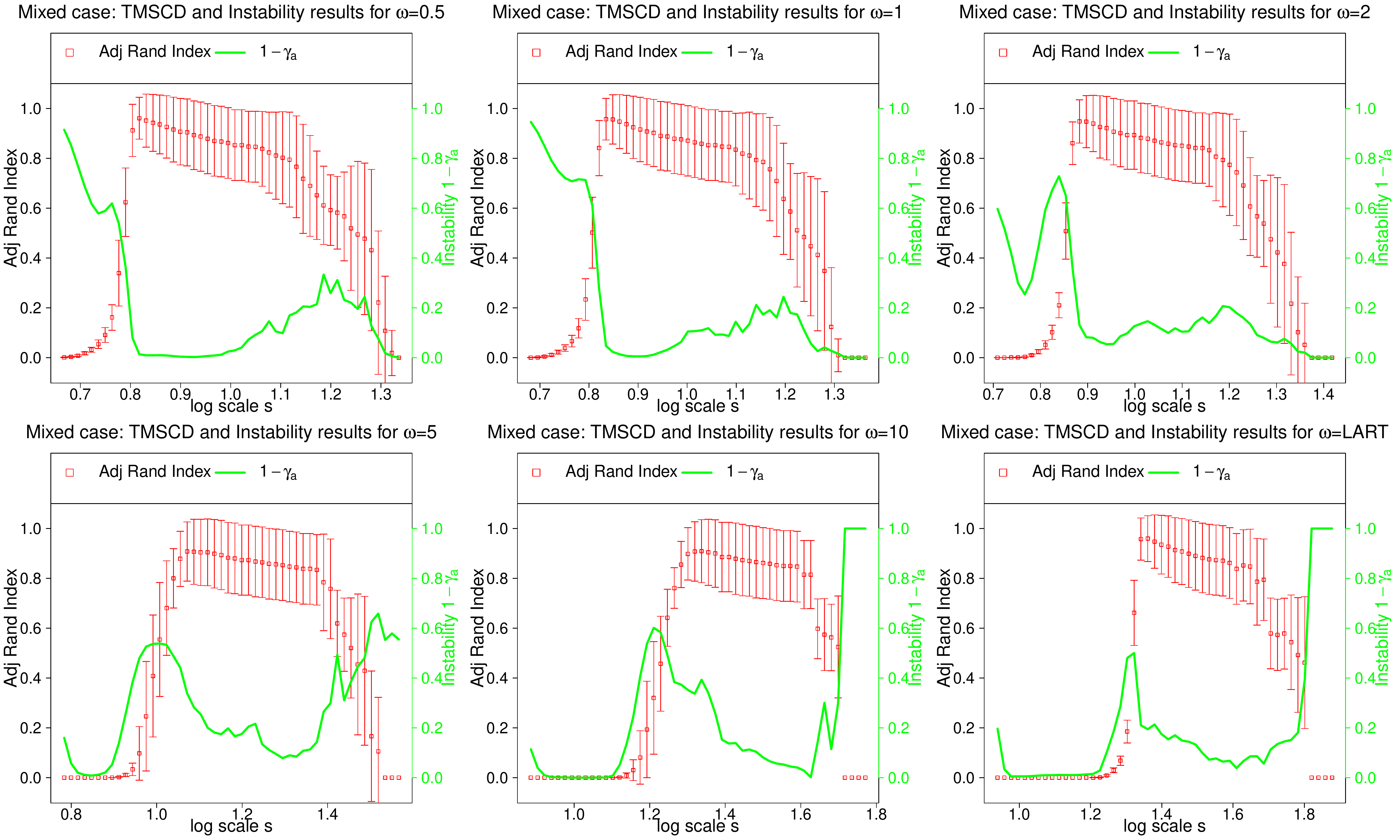}}\\
\subfigure{
\includegraphics[width=\columnwidth]{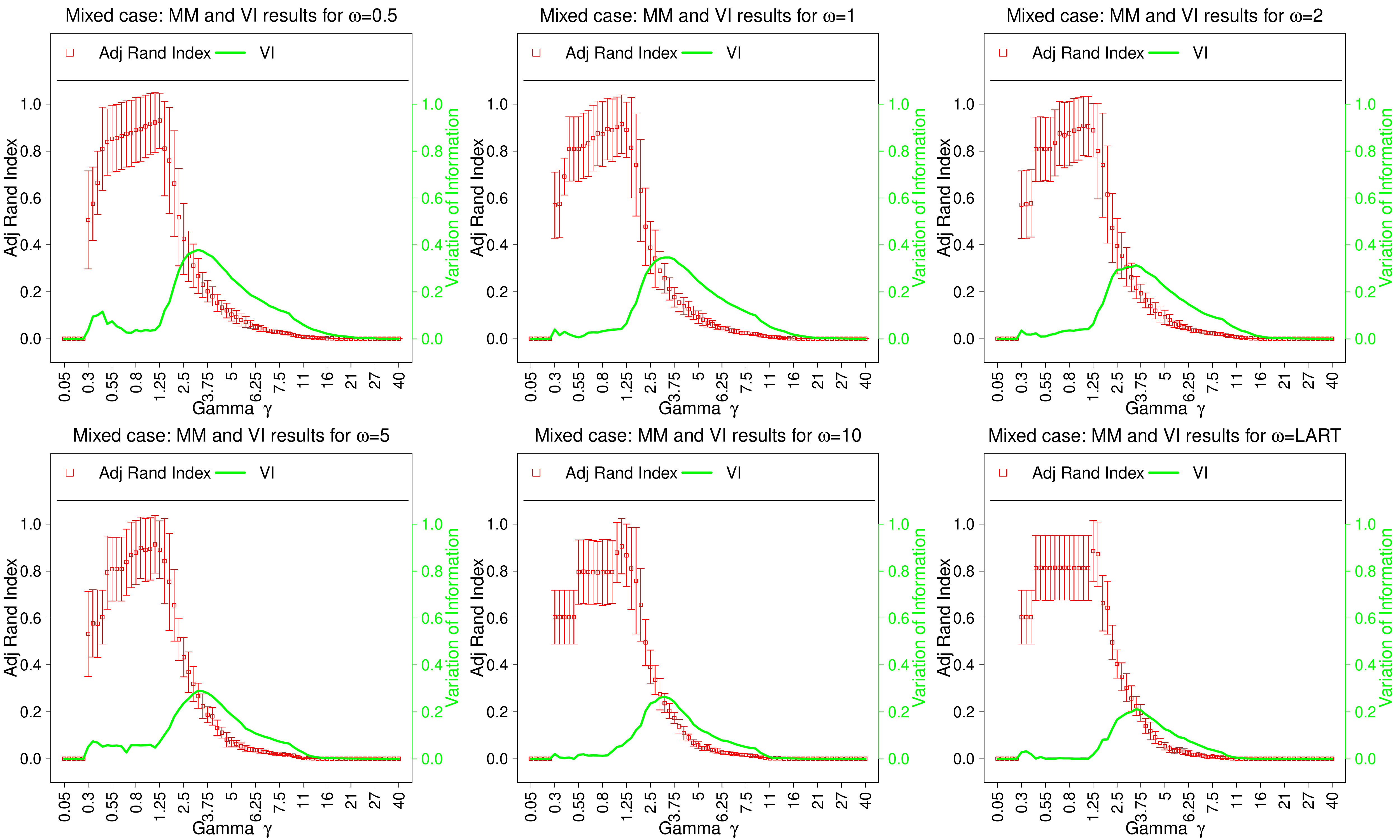}}
\caption[TMSCD and MM results and instability on Mixed benchmark networks.]%
{\textit{TMSCD (top) and MM (bottom) results for Mixed multi-scale benchmark network.} Each pair of plots corresponds to different inter-layer weights $\omega$. First, we plot the results of TMSCD and MM on a realization of Mixed model. Each scale
outputs a partition for all nodes across all time points. For each scale $s$,
we plot the similarity with the theoretical scale, computed as the average over all time
points including std.dev. error bars. In this way, we observe scales where the
exact small (resp. medium, large) scale theoretical partition is uncovered.
Second, for TMSCD we plot the instabilities $1-\gamma_{a}$ versus scale $s$; for MM we
plot variation of information (VI) for each resolution parameter $\gamma$. The
associated partitions of scales with low instability (i.e. high stability)
corresponding to the theoretical partitions. }%
\label{fig:Mixed_Results}%
\end{figure}

For both the TMSCD and MM methods, parameters are set as in ~\ref{sec:MSresults}. The results of TMSCD and MM on a realization of the Grow (resp. Merge, Mixed) network and the instabilities
$1-\gamma_{a}$ versus scale $s$ (VI versus parameter $\gamma$) for different weights
$\omega$ are presented in Figure \ref{fig:Grow_Results} (resp. Figure \ref{fig:Merge_Results} and Figure \ref{fig:Mixed_Results}). 

Overall, there are a couple of observations we can make. First, both methods
perform equally well for the Grow model. In
the Mixed model case, TMSCD performs better than MM and has lower variability in the results. In the Merge model case, TMSCD performs much better than MM
but has larger variability in the results. It appears the Merge model is most
difficult to detect for both MM and TMSCD. This is caused by the nature of the
communities: when two communities are separate they exist at a smaller scale,
but when they merge they exist at a larger scale. Both methods
struggle to perform well for large changes in the sizes of the scales.

Second, the inter-layer weights have effect only over the instability results: TMSCD has more sensitive instability which indicates stable communities
only at the true theoretical partitions.

Third, we observe the appearance of new community scales. For small $s$
and large $\omega$ a new community scale is formed by $N$ communities where
each community is formed by the set of nodes $\left\{  i_{t}%
:t=1,2,...,T\right\}  $ for each $i$. This phenomenon was discussed in the previous Section~\ref{sec:MSresults}. 

We then produce $100$ realizations of each benchmark. Parameters of TMSCD and MM are set as in Section~\ref{sec:MSresults}, where we use LART-type inter-layer weights. The success rate of each realization of a benchmark is the average over the top five adjusted rand index (ARI) values over all scales $s$ or parameter values $\gamma$. The results are summarized in Table~\ref{tab:arenas}.

Both methods perform equally well for the Grow model. In the Mixed model case, TMSCD has better performance with lower variability. In the Merge model case, TMSCD performs much better than MM
but with larger variability in the results. The Merge model is challenging for both MM and TMSCD. This is caused by the nature of the
communities: when two communities are separate they exist at a smaller scale,
but when they merge they exist at a larger scale.
\begin{table}
\centering
\caption{TMSCD and MM results for $50$ realizations of the Grow, Merge and
Mixed model. Each entry is the mean over all realizations $\pm$ one standard deviation. }%
\label{tab:arenas}%
\begin{tabular}
[c]{rccc}
\hline
& Grow & Merge & Mixed\\\hline
TMSCD & $1.0000\pm0.0000$ & $0.8700\pm0.1981$ & $0.9467\pm0.1038$\\
MM & $0.9975\pm0.0088$ & $0.6887\pm0.1302$ & $0.8443\pm0.1412$\\\hline
\end{tabular}
\end{table}
\section{Discussion and Conclusion\label{sec:conclusion}}

The work in this paper is motivated by the need to develop new
methods for multi-scale community detection in temporal networks with automatic
selection of relevant scales. The modularity maximization
\cite{Mucha} achieves excellent results at detecting such communities
but it lacks the flexibility of automatically selecting resolution parameter
ranges relevant to the prevalent community structures over time.
We have used results from Perturbation theory to interpret inter-layer weights as perturbations between time layers, and thus we identify the set of eigenvectors of the supra-Laplacian that are perturbations of the seperate layers' Fiedler vectors. These can be used for detecting communities prevalent over time.

This result gives a completely new point of view on \textit{temporal
networks}. To design the TMSCD method, we reconsidered the role
of the Fiedler vector in community detection for \textit{temporal networks}.
Indeed, the eigenvectors of $\mathcal{L}$ (corresponding to the smallest
eigenvalues) represent all time-layers as separate communities. Hence, we cannot use them for the detection of
communities prevalent over time. We successfully
\textit{attenuated} the influence of these small eigenvalues in the process of
community detection, by properly constructing the wavelet filter function $g$
of the spectral graph wavelets. An important step in our algorithm was the
identification of the uninformative small eigenvalues.

Using simulated data, we have demonstrated that TMSCD
method performs equally well compared to the modularity maximization method
\cite{Mucha}. There are two main advantages to using TMSCD. First, of utmost
importance is the automatic selection of scales at which wavelets should be
obtained and which encompass all relevant within-layer and inter-layer
communities. Second, the proposed LART-type inter-layer weights $\omega_{i}^{t,t+1}$ lead to the
best results in terms of balance between uncovering multi-scale communities and
the stability of those communities at the relevant scales. The stability
procedure used by TMSCD is more sensitive than the modularity
maximization one. This is an advantage when handling real life data sets where
the true scales are not known and a reliable indicator for stable partitions
is required. The supremacy of the LART-type inter-layer weights~\cite{Kuncheva2015} over using
fixed ones indicates the advantage of using adaptable
inter-layer weights that reflect the similarity of nodes across layers.

Given the above results, TMSCD would be an ideal tool for applications to neuroscience and social network analysis.

\section*{Acknowledgments}
ZK acknowledges partial support by the Engineering and Physical Sciences Research Council (EPSRC) and by project DH $02-13$ with Bulgarian NSF.
\bibliographystyle{splncs03}
\bibliography{Lib}

\end{document}